\documentclass[fleqn,usenatbib]{mnras}

\usepackage{newtxtext,newtxmath}

\usepackage[T1]{fontenc}

\DeclareRobustCommand{\VAN}[3]{#2}
\let\VANthebibliography\thebibliography
\def\thebibliography{\DeclareRobustCommand{\VAN}[3]{##3}\VANthebibliography}

\usepackage{graphicx}	
\usepackage{amsmath}	

\newcommand{\herschel}{{\it Herschel}}

\newcommand{\myr}{${\rm M_{\sun}yr^{-1}}$}
\newcommand{\lsun}{${\rm L_{\sun}}$}
\newcommand{\msun}{${\rm M_{\sun}}$}
\newcommand{\mstar}{\ensuremath{M_{\ast}}}
\newcommand{\lya}{Ly$\alpha$}
\newcommand{\fesc}{\ensuremath{f_{\rm esc}({\rm Ly}\alpha)}} 
\newcommand{\lfir}{\ensuremath{L_{\rm FIR}}}

\newcommand{\mic}{$\rm \mu$m}
\newcommand{\off}[1]{} 

\title[Far-infrared observations of dust in LAEs]{Far-infrared observations of dust in \lya\ emitters at $z\approx$\,2\,--\,6}

\author[Rana et al.]{Rahul Rana$^{1, 2}$\thanks{E-mail: rahul.rana@chalmers.se},
Julie L.\ Wardlow$^{2}$,
Thomas\ M.\ Cornish$^{2,3}$, 
Pascale H.\ Desmet$^{2}$,
\newauthor
David Sobral$^{2}$
and Jo\~ao Calhau$^{4}$
\\
$^{1}$Department of Physics and Astronomy, Chalmers University of Technology, SE-412 96, Gothenburg, Sweden\\
$^{2}$Physics Department, Lancaster University, Lancaster, LA1 4YB, UK\\
$^{3}$Department of Physics, Imperial College London, Blackett Laboratory, Prince Consort Road, London SW7 2AZ, UK\\
$^{4}$INAF-Osservatorio Astronomico di Capodimonte, Via Moiariello 16, 80131 Napoli, Italy\\
}

\date{Accepted XXX. Received YYY; in original form ZZZ}

\pubyear{2026}

\begin{document}
\label{firstpage}
\pagerange{\pageref{firstpage}--\pageref{lastpage}}
\maketitle

\begin{abstract} 
The bright \lya\ line is regularly used to identify high-redshift star-forming galaxies known as \lya\ emitters (LAEs). However, \lya\ is affected by resonant scattering and dust absorption making interpretation of its brightness challenging without additional observations. We use SCUBA-2, PACS and SPIRE data to investigate the far-infrared emission, \lya\ escape fraction (\fesc) and infrared excess (${\rm IRX=L_{IR}/L_{UV}}$) in $\sim$4000 LAEs at $z \approx 2.2$~--~6 from SC4K. Five LAEs, all hosting AGN, are individually detected with fluxes $S_{850}=3.7$~--~5.5~mJy at 850~\mic. Stacking is used to probe the average emission from all individually-undetected LAEs, though the stacks are undetected at all wavelengths (e.g.\ $S_{850} < 0.09$\,mJy; $3\sigma$). We group the sample into bins of redshift, stellar mass, \lya\ luminosity and AGN status. Most subsets are undetected but LAEs containing AGN and that have high stellar masses ($\mstar = 10^{10-12}\,{\rm M_{\odot}}$; including and excluding AGN) are detected at most wavelengths, suggesting that stellar mass and AGN heating may be enhancing the dust visibility. Individually detected LAEs and detected stacks have $\fesc=1$--7\%, while all undetected stacks $\gtrsim10\%$. All LAEs together average over $>21\%$ and display significant scatter, suggesting a clumpy ISM dust distribution. Non-zero \fesc\ in massive and AGN-hosting LAEs suggests ionizing photons may escape even from dusty galaxies, challenging the idea that dusty galaxies are poor leakers. Examination of the IRX-$\beta_{\rm UV}$ relation shows LAEs have higher IRX than typical star-forming galaxies at similar redshifts. However, our detections tend to favour more massive, AGN-hosting systems and deeper observations are therefore needed.
\end{abstract}

\begin{keywords}
galaxies: evolution -- galaxies: star formation -- galaxies: high-redshift -- submillimetre: galaxies - galaxies: active
\end{keywords}

\section{Introduction}

In star-forming galaxies, the brightest UV and optical emission line is \lya\ (rest-frame 1216\,\AA), which is produced by the de-excitation of an electron from the $n=2$ to $n=1$ hydrogen energy levels \citep{Partridge1967, Puschnig2020}. It is a powerful tracer of ionized gas around massive, young stars and can reveal channels of low neutral hydrogen through which \lya\ photons escape, providing a rich source of information about the interstellar medium (ISM) and star-forming conditions in galaxies \citep{Partridge1967, Charlot1993, Kunth1998, Hayes2010}. The brightness of \lya\ makes it an effective means of surveying the distant Universe and it is used in both spectroscopic \citep[e.g.][]{Dawson2004, Rhoads2004, Curtis2012, Liu2023, Witstok2025} and narrowband photometric studies \citep[e.g.][]{Tilvi2010, Shibuya2012, Konno2014, Matthee2015, Santos2016, Ota2017, Sobral2018, Hu2019, Wold2022} to efficiently identify high-redshift star-forming galaxies known as \lya\ emitters (LAEs). However, the escape of \lya\ photons from a galaxy is influenced by resonant scattering from neutral hydrogen in the ISM \citep{Neufeld1991, Hansen2006, Smith2022}, the proliferation of the photons to our telescopes are affected by the circumgalactic and intergalactic mediums \citep{Dijkstra2007, Jeeson2012, Mesinger2015, Gurung2020}, and \lya\ is susceptible to absorption by dust grains, since the dust grain size is comparable to the \lya\ wavelength \citep{Atek2008, Hayes2011, Matthee2016}. The \lya\ line can also be energised by AGN \citep{Miley2008}, further complicating the interpretation of the detection of \lya\ from distant systems. Multi-wavelength studies are therefore used to directly characterise LAE counterparts and to infer properties of the subset of the population that may only be detected in \lya\ \citep[e.g.,][]{Nilsson2007, Floranes2012}.

A key parameter governing the interpretation of \lya\ observations is the \lya\ escape fraction, \fesc, defined as the ratio of observed to intrinsic \lya\ emission, which is required to determine the intrinsic properties of LAEs from the detected \lya\ flux. \fesc\ can be ascertained by using other measurements to predict the intrinsic \lya\ luminosity. For instance, the \lya\ to H$\alpha$ ratio is expected to be $\sim8.7$ under case B recombination conditions \citep{BakerMenzel1938}, so a  lower observed ratio indicates the scattering or attenuation of \lya\ photons. This method has been used in a number of studies \citep[e.g.][]{Sobral2018, Matthee2021, Melinder2023, Ning2023, Roy2023}, including to show that LAEs at $z \approx 2$ have $\fesc\sim 2$~--~30\%, with a median of $\sim$1.6\%, which decreases for galaxies with higher SFRs and dust attenuation  \citep{Matthee2016}. However, in the presence of AGN the intrinsic Ly$\alpha$/H$\alpha$ ratio departs from simple Case~B (because AGN ionizing spectra, shocks, and complex radiative transfer alter line production and escape), making Ly$\alpha$/H$\alpha$ an unreliable estimator of the \fesc\ without decomposition into AGN and star-formation components \citep[e.g.,][]{Matthee2016}. Furthermore, the \lya\ to H$\beta$ ratio is predicted to be $\sim$23 under case B recombination, so this ratio can be similarly employed to determine \fesc\ \citep[e.g.][]{Erb2019, Saxena2023}. Using this method \citet{Weiss2021} observed mean $\fesc\sim$6\% at $z\sim2$, with values decreasing with increasing stellar mass. However, if these galaxies host AGN, Ly$\alpha$/H$\beta$ can also be strongly affected by nuclear emission and scattering, leading to large scatter and unreliable \fesc\ estimates \citep[e.g.,][]{Scarlata2009}. There is also an observed correlation between observed \lya\ equivalent width (EW) and \fesc, which can be used to infer \fesc\ \citep[e.g.][]{Verhamme2017, Harikane2018, Sobral2019}. 

However, in practise these techniques have further limitations: dust attenuation affects the derivation of \fesc\ from UV and optical observations (e.g.\ H$\alpha$, H$\beta$ and \lya\ lines; \citealt{Vale2020}), the H$\beta$ line is intrinsically faint and difficult to detect, and the correlation between \lya\ EW and \fesc\ is only calibrated for $z \leq 2.6$ and ${\rm EW}\leq 160$\,\AA. To overcome these shortcomings \citet{Calhau2020} stacked  $\sim3700$  LAEs at $z\sim2$--6 in the radio to derive a dust-free SFR with which to estimate the intrinsic \lya\ output, and finding a median $\fesc = 50\pm20\%$ from the ratio of observed to intrinsic \lya. However, the derivation of \fesc\ from radio emission depends on the calibration between radio continuum and SFR, which can be significantly affected by AGN contamination  \citep{Bourne2011, Delhaize2017, Daniel2018}. Thus, measurements of \fesc\ in LAEs remain uncertain and additional comparison methods are required.  

Since \lya\ is predominantly generated in star-forming regions, an alternative is to calculate the total star-formation rate of a galaxy and use this to predict the intrinsic \lya\ luminosity in the calculation of \fesc\ \citep[e.g.][]{Wardlow2014}. Far-infrared emission provides insight into the dust-obscured SFR and combining this with UV-based SFR measurements provides a total SFR \citep{Wuyts2011, Oteo2014, Boquien2016}. These measurements complement each other and eliminate the need to apply dust attenuation corrections to UV-based SFR, which rely on assumptions that are not yet well-constrained for LAEs. One challenge with using far-infrared emission to probe \fesc\ in LAEs is that these galaxies were expected to be dust-poor because the presence of even small amounts of dust can attenuate the \lya\ to below detectable levels. Furthermore, the presence of high EW \lya\ lines implies young stellar populations, that have not had enough time to enrich the ISM with significant amounts of dust through supernovae or AGB evolutionary phases \citep[e.g.][]{Finkelstein2009}. However, reddening of the UV/optical SEDs of LAEs suggests that old stars or dust may be present \cite[e.g.][]{Chary2005, Finkelstein2009, Shimizu2011}. In this scenario, clumpy dust can completely absorb the \lya\ on some lines of sight, but others can be completely free from dust and with the photons able to escape. Indeed, for intermediate to high-redshift galaxy populations it is increasingly being shown that dust and \lya\ emission can co-exist especially if the dust is clumpy or unevenly distributed \citep[e.g.][]{Haiman1999, Scarlata2009, Oteo2012}. In addition, It has been demonstrated that dusty star-forming galaxies (DSFGs) can exhibit blue ultraviolet spectral slope ($\beta_{\rm UV}$), which may lead to an underestimation of the obscured star formation rate \citep[e.g.,][]{Casey2014b, Casey2026}. Therefore, to accurately estimate the dust correction and total SFR it is important to account for regions that are completely optically-thick at UV and optical wavelengths. 

Until now, the limited availability of deep far-infrared data has led to a dearth of studies targeting far-infrared emission from LAEs. Single-dish far-infrared and submillimetre telescopes offer wide-field imaging capabilities but are hampered by high confusion limits due to their large beam sizes, e.g.\ the \textit{Herschel}/SPIRE beam is $\sim18^{\prime\prime}$ at 250\,\mic\ \citep{Griffin2010} and at 850\,\mic\ James Clerk Maxwell Telescope (JCMT)/SCUBA-2 has a $\sim14^{\prime\prime}$ beam \citep{Holland2013}. Interferometers like ALMA can deliver deeper data, but have narrow fields-of-view, which makes it observationally expensive to survey large samples of faint sources such as LAEs \citep[e.g.,][]{Ota2014, Williams2014}. Instead, stacking can be used. For example, \citet{Wardlow2014} attempted to detect dust emission to measure \fesc\ from $\sim500$ LAEs at $z=2.8$--4.5 using \textit{Herschel} and LABOCA data. They did not detect any individual galaxies and stacks of samples in three redshift bins were reached a maximum of $2.8\sigma$, leading to limits of $\fesc \gtrsim10\%$: slightly higher than the global \fesc\ evolution at the targeted redshifts. Similarly, \citet{Haruka2015} stacked \textit{Spitzer}/MIPS and \textit{Herschel}/PACS observations of 213 LAEs at $z\simeq2.18$ in GOODS-South, leading to an estimate of the  average \fesc\ of 16--37\%, which is also higher than the cosmic average at the same epoch.

Direct observations of infrared emissions are also used to correct for the impact of dust attenuation on galaxy parameters that are derived from rest-frame UV and optical measurements (e.g.\ UV luminosities, star formation rates, stellar masses). 
Once constrained, the empirical relationship between the infrared excess (IRX=$\rm{L_{IR}/L_{UV}}$) and $\beta_{\rm UV}$; (see e.g.\ \citealt{Meurer1999, Calzetti2000, Gordon2003, Reddy2018}) can be used to perform such corrections even for galaxies that are not individually detected in IR observations \citep[e.g.,][]{Battisti2016}. However, at high redshifts the IRX-$\beta_{\rm UV}$ relation has mainly been tested for star-forming or Lyman-break galaxies (LBGs) (e.g.\ \citealt{Capak2015, Alvarez2016, Bouwens2016, Barisic2017, Bowler2018, Koprowski2018, Fudamoto2020}) with few constraints for LAEs, due to the challenges in detecting dust emissions from LAEs. For example, \citet{Haruka2015} stacked {\it Spitzer}/MIPS observations of LAEs at $z\simeq 2.18$ to constrain ${\rm IRX}\le 2.2$ for $\beta_{\rm UV}=-1.4\pm0.2$, consistent with the SMC and local starburst extinction curves \citep{Pettini1998, Meurer1999, Takeuchi2012}, though detection of individual LAEs and samples at different redshifts are still required.  

In this study, we use  100, 160, 250, 350, 500 and 850\,$\mu$m continuum observations from \herschel/PACS, \herschel/SPIRE and JCMT/SCUBA-2 to investigate the dust emission from $\sim4000$ LAEs at $z= 2.2$--6 from the SC4K survey \citep{Sobral2018} of the COSMOS field. We search for dust emission from individual LAEs and use stacking analyses to measure the average infrared fluxes from LAEs that are too faint to be individually detected. These results are used to probe \fesc\ and the IRX-$\beta_{\rm UV}$ relation for LAEs. The paper is organized as follows: in Section~\ref{Sec:2}, we discuss the LAE sample and the far-infrared data. Section~\ref{Sec:3} outlines the flux measurement and stacking methods employed in this analysis. Spectral energy distributions (SEDs), escape fractions and the IRX-$\beta_{\rm UV}$ relation are discussed in Section~\ref{Sec:4} and our conclusions are presented in Section~\ref{Sec:5}. Throughout this work, we adopt $\Lambda$CDM cosmology with $H_{0} = 69.3\, {\rm kms^{-1}\,Mpc^{-1}}$, $\Omega_{\rm M} = 0.29$, and $\Omega_{\Lambda} = 0.71$ \citep{Hinshaw2013} and apply a Chabrier \citep{Chabrier2003} initial mass function (IMF).

\section{Data} 
\label{Sec:2}

Since LAEs are expected to be faint at far-infrared wavelengths this work requires a large sample of LAEs in an area with deep submillimeter observations. We therefore focus on the COSMOS field. 

\subsection{LAE Sample}\label{Sec:LAE_sample} 

We begin by considering the 3908 LAEs identified by SC4K \citep{Sobral2018} in deep medium- and narrow-band observations of $\sim 2\,{\rm deg}^2$ of COSMOS. 
The SC4K survey used imaging from 12 medium-band filters available in the COSMOS archive \citep{Ilbert2009,Taniguchi2015} obtained using Suprime-Cam mounted on the Subaru Telescope \citep{Miyazaki2002}, with limiting magnitudes ranging from 25.4 to 26.2\,AB\,mag. The survey used narrow-band imaging from four narrow-band surveys, including those presented by \citet{Sobral2017b} and \citet{Matthee2017b}, which were conducted with the 2.5\,m Isaac Newton Telescope and achieved a $5\sigma$ depth of 24.0\,AB\,mag, as well as observations in narrow band filters obtained with Suprime-Cam \citep{Santos2016}. SC4K identified LAEs at $z \sim 2.2$\,--\,6 with observed \lya\ EW $>50 (1 + z)$\,{\AA} in 16 redshift slices to trace the evolution of typical LAEs and the \fesc\ from $z \sim 2.2$ to 6 over a comoving volume of $\sim10^{8}$\,Mpc$^3$. SC4K selected LAEs based on a clear narrow-band colour excess followed by identifying candidate LAEs using photometric redshifts, Lyman-break and colour cuts, and by removing lower-redshift line contaminants. After visual inspection, they found 3434 medium-band LAEs and 474 narrow-band LAEs, giving a final sample of 3908 LAEs. \lya\ luminosities for these LAEs were computed using \lya\ fluxes in 2$''$ aperture and luminosity distances based on \lya\ line which correspond to $\sim10^{42.4}$\,erg s$^{-1}$ at $z\approx2.5$ and $10^{43.0}$\,erg s$^{-1}$ at $z=$\,5.8.

Our analyses require parameters derived from spectral energy distribution (SED) fitting, which was performed by \citet{Santos2020} using high-redshift extension of {\sc magphys} \citep{daCunha2008}. They assumed a Chabrier IMF and the  \citet{Charlot2000} dust attenuation model and derived physical parameters from posterior probability distributions. From the best-fitting SEDs they computed $M_{\mathrm{UV}}$ by integrating over 1400--1600 \AA, spanning $-23 \lesssim M_{\mathrm{UV}} \lesssim -17$. In addition, they derived stellar masses of SC4K LAEs and found that they are mostly below $10^{10}$ \msun\ with a median $M_\star \approx 10^{9.3}$ \msun\ with a mild redshift-dependent increase due to selection effects. The UV slope $f_\lambda \propto \lambda^\beta$ is inferred from the SEDs, but {\sc magphys} imposes a hard limit of $\beta = -2.44$, restricting how blue the models can be. \citet{Santos2020} flagged LAEs lacking sufficient multi-band photometry for reliable SED fitting as sources with ``bad" SEDs. Furthermore, SC4K AGN LAEs identified by \citet{Calhau2020} through X-ray (\textit{Chandra}) and/or radio (VLA) counterparts, using positional cross-matching and luminosity-based criteria to isolate X-ray– and radio-bright sources indicative of accretion-driven activity were fitted without using any AGN-specific SED model. We exclude the 225 SC4K LAEs that are flagged as having ``bad" SED fits, leaving a sample of 3683 LAEs that are studied here. 

\begin{table*}
\caption{Multi-wavelength (100–850\,\mic) flux densities of LAEs individually detected in the far-infrared data.}
\label{tab:six_LAEs_flux1}
\begin{tabular}{lccccccc} 
\hline
Source ID$^{\psi}$  & $z$ & $S_{100}$$^a$ &$S_{160}$$^a$ & $S_{250}$$^a$ & $S_{350}$$^a$ & $S_{500}$$^a$ & $S_{850}$$^a$ \\
 & & (mJy) & (mJy) & (mJy) & (mJy) & (mJy) & (mJy) \\
\hline
SC4K-IA505-178627 & 3.0837±0.0003$^{e}$ & 1.2±0.5 & 3.6±1.2 & 7.0±1.6   & 10.4±2.8 & 11.1±3.2 & 3.7±1.0 \\
SC4K-IA574-34828  & 3.61±0.11       & 7.3±3.0 & 10.5±4.3    & 17.1±3.6 & – $^g$        & 15.6±8.3 & 5.5±1.4 \\
SC4K-IA484-28746  & 2.98±0.10       & 3.1±1.8 & 16.7±2.2    & 18.5±3.5 & 0±8       & 6.1±4.8 & 3.7±0.9 \\
SC4K-IA484-69327  & 2.92±0.10       & 6.4±2.1 & 13.4±9.5    & 1±5$^f$ & 34.0±10.3  & 21.2±7.4 & 4.5±1.2 \\
SC4K-IA679-223923 & 4.629±0.005$^{e}$& 2.4±1.7 & 14.3±4.1   & 7.4±1.8 & 12.4±2.1   & 16.1±3.4 & 3.9±1.2 \\
\hline
\end{tabular}
\\
$^{\psi}$ Source IDs are from SC4K \citep{Sobral2018}. 
$^a$ Flux densities at 100--500\,\mic\ are from the \herschel\ PACS and SPIRE super-deblended catalogue \citep{Jin2018} and for 850\,\mic\ are from the deblended \citet{Simpson2019} catalogue.
$^e$ LAEs with spectroscopic redshifts from \citet{DESI_Collaboration2024} and \citet{Hasinger2018}; all other redshifts are from the detection of \lya\ in medium- and narrow-band photometry \citep{Sobral2018}.
$^f$ The value is excluded from SED fitting due to being inconsistent with the other photometry for the same source (Section~\ref{sec:sedfit}).
$^g$ Reliable flux-density measurement is not available for the source \citep{Jin2018}.
\end{table*}

\subsection{Far-infrared data}
\label{sec:firdata}

\subsubsection{\textit{Herschel} PACS and SPIRE}
\label{Sec:SPIRE}

The whole COSMOS field was mapped by \herschel\ with both the PACS and SPIRE instruments. In this study we use \textit{Herschel}/PACS 100 and 160\,\mic\ data from the public release of the PACS Evolutionary Probe (PEP; \citealt{Lutz2011}) and  \textit{Herschel}/SPIRE 250, 350 and 500\,\mic\ data from the Herschel Multi-tiered Extragalactic Survey (HerMES; \citealt{Oliver2012}) that has been reprocessed by the Herschel Extragalactic Legacy Project (HELP; \citealt{Shirley2021}). The PACS maps have pixel scales of 1.2 and 2.4\,arcsec and point spread functions (PSFs) of 7.2 and 12\,arcsec, respectively, and reach depths of $\sigma_{100}$ = 1.50 and $\sigma_{160}$ = 3.27 mJy. The SPIRE maps have pixel scales of 6, 8.3 and 12\,arcsec and beams of 18.15, 25.15, and 36.30\,arcsec at 250, 350 and 500\,\mic, respectively.
The noise levels are $\sigma_{250} = 3.18$, $\sigma_{350} = 2.66$ and $\sigma_{500} = 3.82$\,mJy in COSMOS \citep{Oliver2012, Shirley2021}.  

Examination of the distribution of pixel values show that the median values of the SPIRE maps are slightly negatively offset, which would artificially decrease flux densities derived from stacking unless corrected. These maps are produced from standard pipeline reductions that remove baseline offsets; as a result pixel-value distribution can peak near or slightly below zero \citep{Hurley2017, Shirley2021}. We thus correct flux measurement by adding the flux offsets of 1.62, 1.65 and 1.3 mJy to each pixel of 250, 350 and 500 \mic\ maps respectively. 

\subsubsection{SCUBA-2 850 \mic} 
\label{Sec:SCUBA-2}
To probe longer wavelengths we use JCMT/SCUBA-2 850\,\mic\ data from S2COSMOS \citep{Simpson2019}, which reaches a median noise level of $\sigma_{850} = 1.2$\,mJy and has a pixel scale of 2\,arcsec. The 14.6\,arcsec resolution map is centred on the COSMOS field and covers $\sim2.6\,{\rm deg}^2$. 

The standard S2COSMOS map has been processed with a matched filter, which helps to detect individual sources, but introduces negative rings around bright sources \citep{Simpson2019}. Thus, LAEs that are situated in the negative regions in the match-filtered map would systematically add negative flux when stacking. To prevent these negative rings in the match-filtered map from artificially reducing any stack signals, we instead work from a version of the S2COSMOS map that has not been match filtered \citep{Simpson2019}, which we process using 2D Gaussians to replicate the S2COSMOS match-filtering process but without the negative PSF wings. The map is first convolved with a 30\,arcsec 2D Gaussian and this convolved map is subtracted from the original in order to filter out large-scale residual noise.  We then convolve the resulting map with a 9.6\,arcsec 2D Gaussian such that the resulting image has the same 14.6\,arcsec PSF as the instrument and the S2COSMOS match-filtered map.  Finally, a flux conversion factor (FCF) is estimated by comparing the flux densities of bright sources in our Gaussian-smoothed map with the same sources in the match-filtered images. Applying this FCF leads to our final corrected map, which has the same PSF and source flux densities as the S2COSMOS match-filtered map but without the negative rings around bright sources.

\begin{figure*}
\begin{center}
\includegraphics[scale=0.5]{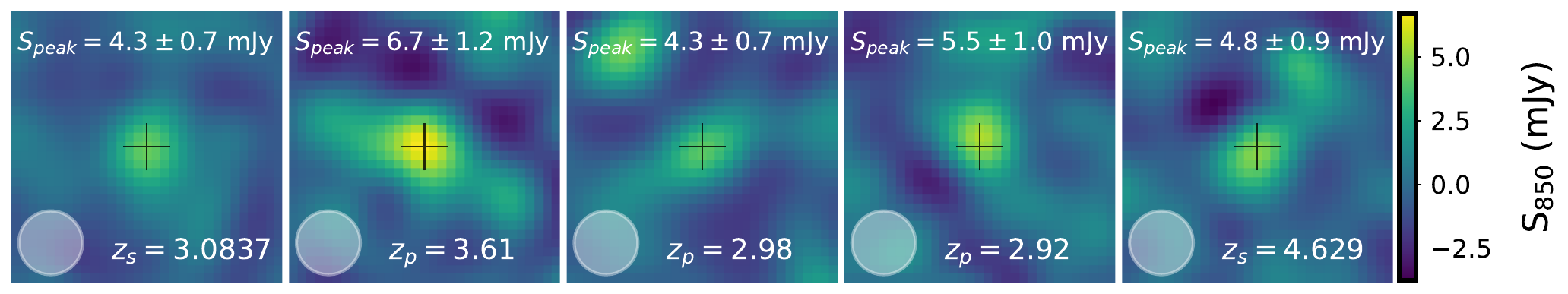}
\caption{$62\times62$\,arcsec SCUBA-2 850\,$\mu$m cutouts at the positions of the five individually detected LAEs with the redshifts ($z_s$ are spectroscopic redshifts and $z_p$ are photometric redshifts) and peak flux densities of each SCUBA-2 detection labelled. LAE positions are marked by black crosses and white circles indicate the size of the 14.6\,arcsec SCUBA-2 beam.} 
\label{fig:six_LAEs}
\end{center}
\end{figure*}

\section{Flux Measurements} \label{Sec:3}

\subsection{Individual detections}
\label{sec:indivduals}
To examine dust properties of individual LAEs, we begin by determining whether any of the SC4K LAEs are individually detected in the S2COSMOS data. To accomplish this, we cross-match the positions of SC4K sources to the S2COSMOS catalogue \citep{Simpson2019} using a 7\,arcsec matching radius (i.e.\ the radius of the SCUBA-2 850\,\mic\ beam) and find 31 matches. RR and JLW independently visually inspected these 31 sources using archival imaging from \textit{Hubble} \citep[e.g.][]{Scoville2007}, Subaru \citep{Taniguchi2015}, UKIRT \citep{Lawrence2007},  \textit{Spitzer} \citep{Sanders2007}. We retained only those LAEs for which both inspectors reached the same conclusion, identifying systems in which the multi-wavelength data support an association between the 850\,\mic\ emission and the LAE position. Higher-resolution optical and radio data, along with possible infrared counterparts in \textit{Spitzer} and/or UKIRT observations helped us identify LAEs with far-infrared emission. Of the 31 individual matches, we confirm five individual LAEs are detected in S2COSMOS, and there are an additional four LAEs that likely blended with nearby sources in 850\,\mic\ map. The five LAEs with 850\,\mic\ emission are all within 3\,arcsec of the corresponding S2COSMOS source (Figure~\ref{fig:six_LAEs}) and the four blended LAEs are matched to within 4.5\,arcsec.

\begin{figure}
\begin{center}
\includegraphics[scale=0.37]{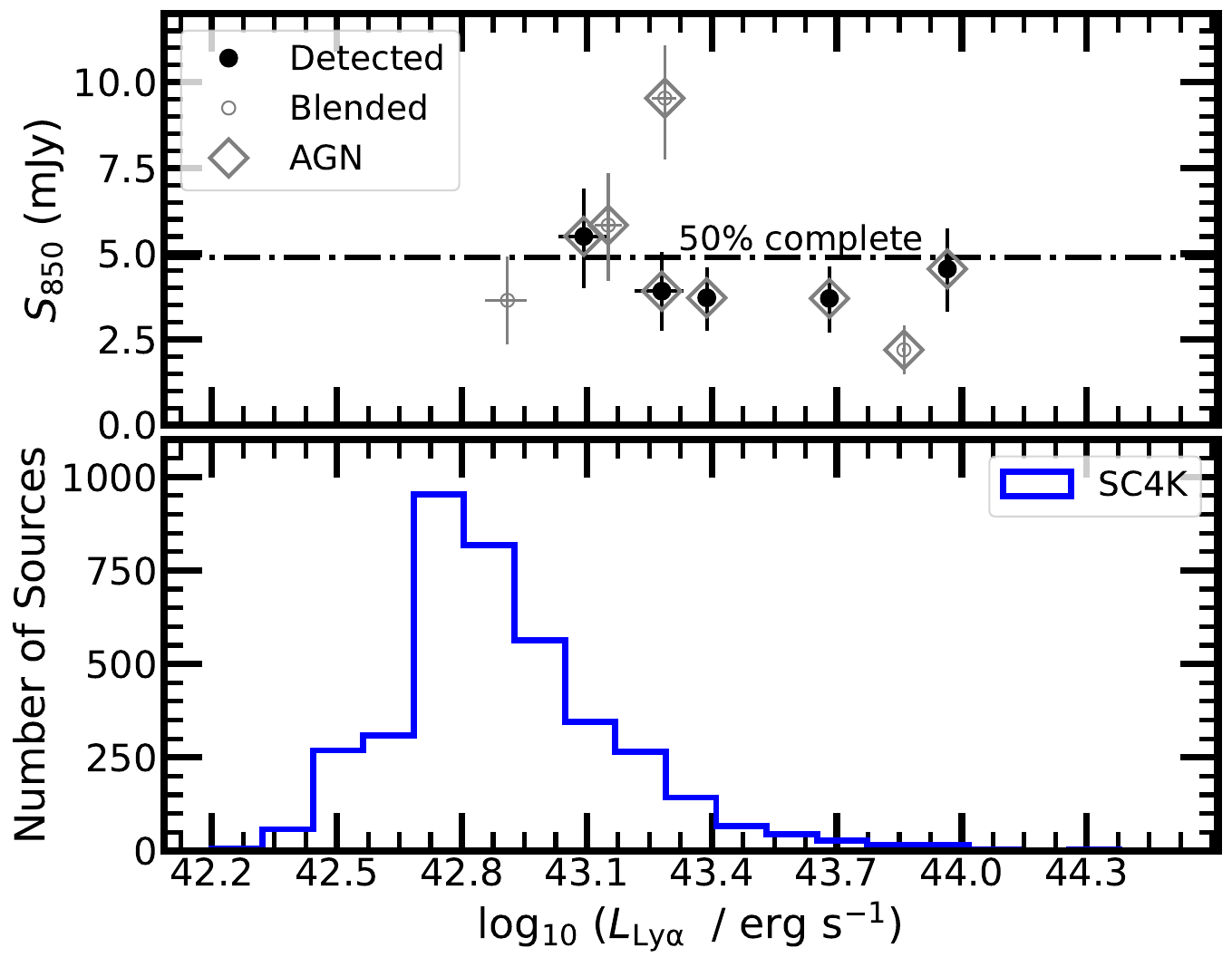}
\caption{850\,$\mu$m flux densities and \lya\ luminosities of the five  LAEs that are individually-detected in the submillimetre, compared with the S2COSMOS 50\% completeness limit (\textit{top panel}) and the distribution of \lya\ luminosities of the whole SC4K catalogue (\textit{bottom panel}). The four blended sources are also shown and those with X-ray and/or radio AGN activity are highlighted. The five individually detected and four blended LAEs in S2COSMOS lie at the high–\lya-luminosity end of the population and host AGN. Consequently, this small sample of S2COSMOS-detected LAEs appears to be biased toward higher \lya\ luminosities and a higher AGN fraction compared to the full LAE population, which has an AGN fraction of 8$\%$, although the limited sample size limits statistical significance.} 
\label{fig:flux_comparision}
\end{center}
\end{figure}

The five individually 850-\mic\ detected LAEs are shown in Figure~\ref{fig:six_LAEs} and Table~\ref{tab:six_LAEs_flux1}. They are detected at 5.1--$6.2\sigma$ and are at $z\sim2.92$\,--\,4.63 and we take their 850\,\mic\ flux densities from the S2COSMOS catalogue \citep{Simpson2019}, which accounts for deboosting. For these five LAEs we use the LAE positions to extract \herschel\ PACS and SPIRE fluxes from the publicly accessible super-deblended COSMOS catalogue \citep{Jin2018}, wherein the far-infrared emission has been deblended to align with optical-NIR coordinates. All five of these LAEs that are bright enough to be individually-detected in submillimetre data are X-ray and/or radio-detected AGN hosts \citep{Calhau2020} with Ly$\alpha$ luminosities $> 10^{42.8}$\,erg\,s$^{-1}$ (SC4K median). Figure~\ref{fig:flux_comparision} shows that both the individual detections and three of four likely blended LAEs contain X-ray and/or radio AGN and that they are more luminous in \lya\ than the median of the SC4K sample as a whole. The SC4K sample contains $\sim8\%$ AGN, while all five of our detections host AGN. It is therefore possible that the \lya\ and/or infrared emission from these systems may include a contribution from the AGN, though this is uncertain (see e.g.\ \citealt{Calhau2020}). 

\subsection{Stacking} \label{Sec:3.2}

Since most LAEs are not bright enough in the far-IR to be individually detected in the SCUBA-2 and \herschel\ observations, we next use stacking to probe the average dust emission from this population. 
Stacking a number of sources, $N$, reduces background noise by a factor of  $\sqrt{N}$, so stacking the whole population of $\sim4000$ SC4K LAEs enables the measurement of emission $\sim60\times$ deeper than is possible for individual systems.  We exclude the five individually detected and four blended LAEs (Section~\ref{sec:indivduals}) from all stacks so as to avoid these potential outliers from biasing the results.

We write a bespoke {\sc Python} stacking code, which produces cutouts of the  PACS, SPIRE and SCUBA-2 maps described in Section~\ref{sec:firdata}  at the positions of the SC4K LAEs. These are then median combined to determine the median flux of the LAEs that are included in each stack. The size of the stacks is scaled to the beam size at each wavelength, such that we produce median stacks that are $\sim 37^{\prime\prime}\times37^{\prime\prime}$,  $74^{\prime\prime}\times74^{\prime\prime}$, $150^{\prime\prime}\times150^{\prime\prime}$, $208^{\prime\prime}\times208^{\prime\prime}$, $300^{\prime\prime}\times300^{\prime\prime}$ and $62^{\prime\prime}\times62^{\prime\prime}$ for the 100, 160, 250, 350, 500 and 850\,\mic\ median stacks, respectively. 

To verify the astrometric alignment between the far-infrared datasets we first stack the \herschel\ and SCUBA-2 maps at the positions of publicly available \textit{Spitzer}/MIPS \citep{Frayer2009} and VLA 3\,GHz \citep{Smolcic2017} selected sources, which are expected to be bright in the far-infrared \citep[e.g.,][]{Ivison2008, Lee2010, Biggs2011}.  These {\it Spitzer} and VLA stacks reveal robustly detected signals that are centred in the cutouts in all six \herschel\ and SCUBA-2 datasets, which demonstrates the power of stacking and confirms excellent astrometric alignment between the datasets, consistent with the residual astrometric offset of $\approx0.2^{\prime\prime}$ for the SCUBA-2 850\,\mic\ \citep{Simpson2019}, while the Herschel data astrometry is limited by the telescope pointing error of $\leq1^{\prime\prime}$ \citep{Sanchez2014}.

\begin{figure*}
\begin{center}
\includegraphics[scale=0.86]{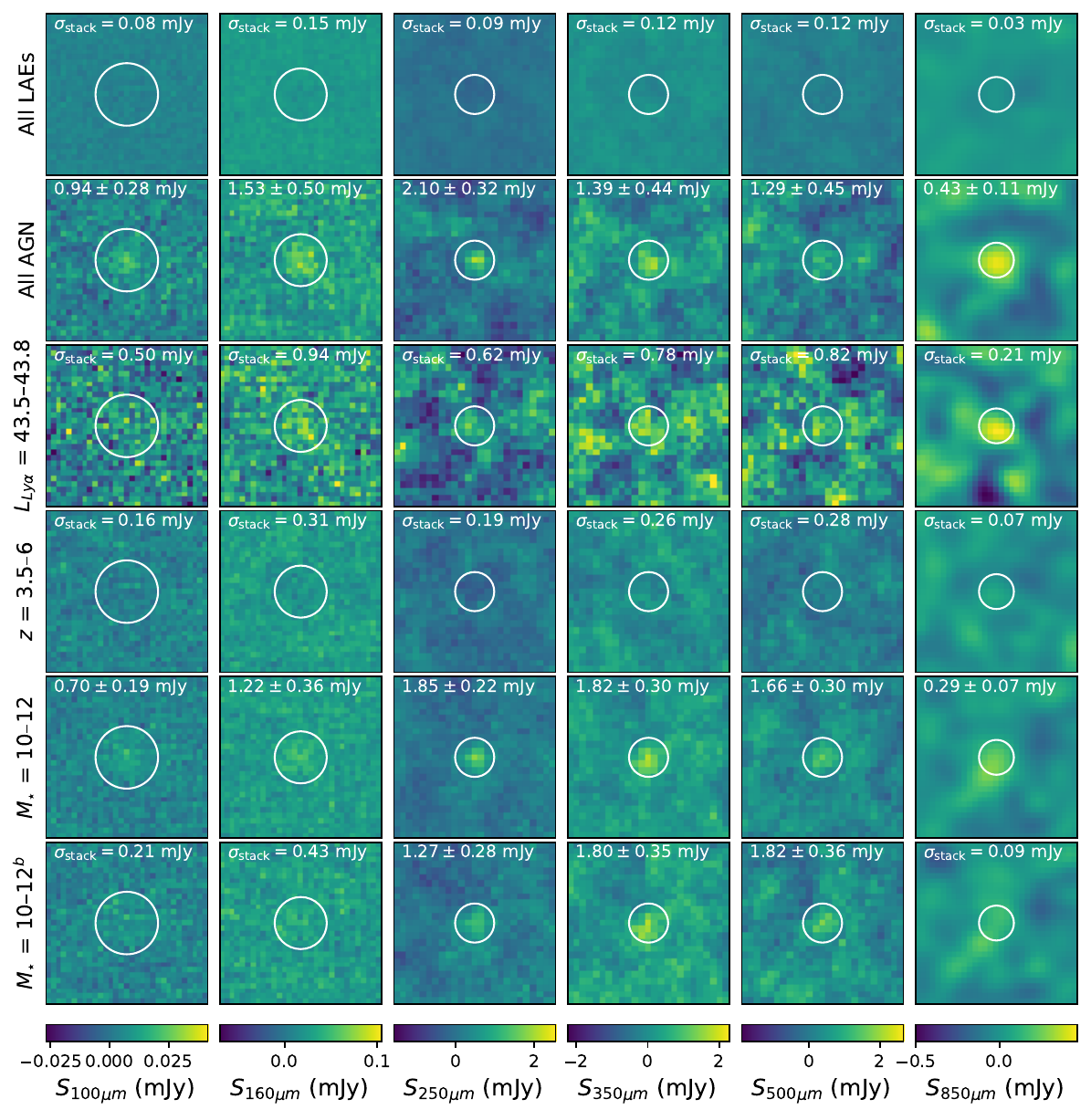}
\caption{Images of six of our median stacks. Rows correspond to different subsets (labelled on the left) and columns show the flux densities at different wavelengths (labelled with colour bars at the bottom). From top to bottom, the example stacks shown are all LAEs, all LAEs that contain AGN, LAEs with $\log_{10}({\rm L_{Ly\alpha} / erg\,s^{-1}} ) = 43.5$\,--\,43.8, those with $\log({\rm M_{\star} / M_{\odot}}) = 10$--12 and those in the same mass bin but with AGN excluded (denoted by $^b$), and those at $z=3.5$--6. Colour bars are showing flux densities in mJy/pixel for PACS (100 and 160 \mic) and in mJy/beam for the other stacks. White circles are located at the centres of the stacks and represent the beam sizes at the relevant wavelength. The sizes of the cutouts are discussed in Section~\ref{Sec:3.2}. For subsets that are detected at $\ge3\sigma$ the labels give the signal and noise values; for samples that are predominantly not detected in our stacks the background noise level is instead given.}
\label{fig:stacks}
\end{center}
\end{figure*}

\subsubsection{Stacked photometry}
The \herschel/PACS maps are in units of Jy/pixel and we therefore use PSF aperture photometry as described in \citet{Santos2020} and \citet{Lutz2011} and centred at the centre of each stack.  
Apertures that match the PSF size i.e.\ 7.2 and 12\,arcsec at 100 and 160\,\mic, respectively are used, with background annuli extending from 2 to $3\times$ further than this. To ensure accurate flux measurements, correction factors are applied following the procedures recommended in \citet{Balog2014}. Specifically, an aperture correction factor of 1.49 is used for both measurements to account for flux losses outside the chosen aperture. In addition, filter-dependent correction factors of 1.2 for the 100~$\mu$m band and 1.1 for the 160~$\mu$m band are applied. These values are adopted based on the calibration and methodology outlined in \citet{Popesso2012} and \citet{Balog2014}.

SPIRE maps are instead in units of mJy/beam, wherein for unresolved sources (as expected for high redshift galaxies), the brightest pixel gives the total flux. Therefore, to determine the flux from 250, 350 and 500\,\mic\ stacks, we verify that the brightest pixel is at the centre of the maps and use the value of this pixel as the stacked flux. 

The S2COSMOS maps are also in units of mJy/beam and therefore the brightest pixel contains all the flux for a point source. However, in this case the brightest pixel is not always the central pixel -- likely due to small spatial offset between the \lya\ and dust emission regions combined with SCUBA-2 pixel scale and beam size. We visually inspect the individual LAE cutouts that contribute to the stack and verify that the offsets are not driven by a bias in a directional offset of some bright sources, but are instead due to randomisation amongst all LAEs. Extracting robust flux densities from the SCUBA-2 maps requires first identifying the brightest pixel within the beam associated with the central stacked pseudo-source i.e.\ excluding noisy pixels near the edges of the beam. For this we use a two step process to identify the brightest pixel within a SCUBA-2 beam size at the centre of the stack that is driven by emission from within (and not outside) this region. This two step process is required to ensure that the outskirts of non-central noise peaks are not inadvertently selected in cases where there is no significant LAE detection. We first identify the brightest pixel ($p_1$) within the beam around the centre of the stack; a second circle of beam radius is then drawn around $p_1$, and the brightest pixel within this second circle, denoted as $p_2$, is determined. If $p_2$ coincides with the centre of the second circle, it is likely associated with the stack and is used to measure stack flux density. Conversely, if $p_2$ is offset from the centre, it is likely associated with a nearby source/noise, and the flux density of the stack is instead measured using $p_1$, though in this case the stack is likely undetected (i.e.\ ${\rm SNR}<3$). This procedure is repeated in the process of measuring photometric uncertainties. 

To calculate the uncertainty in the flux density of the far-infrared stacks we select random positions in the SC4K coverage region, with masking to avoid bright stars and artefacts. The number of random positions used is set equal to the number of LAEs included in each stack (Table~\ref{tab:Stack_flux}). We then stack the far-infrared data at these random positions and measure the flux densities for each dataset as described above. This random stacking process is repeated 1000 times for each stacked dataset and the standard deviation of the flux density of these 1000 random stacks provides the uncertainty on the corresponding LAE stack. We have confirmed that the medians of these distributions are consistent with 0\,mJy, as expected for random positions. 

\begin{table*}
\centering
\setlength{\tabcolsep}{5.2pt}
\caption{100--850\,\mic\ flux densities measured for stacks of different subsets of SC4K LAEs. Values corresponding to $\geq 3\sigma$ detections are shown in bold.}
\label{tab:Stack_flux}

\begin{tabular}{lcccccccc}
\hline

Sample & \# Sources$^a$ & $z_{\rm median}$$^c$ & $S_{100}$ & $S_{160}$ & $S_{250}$ & $S_{350}$ & $S_{500}$ & $S_{850}$ \\
\hline

All LAEs   & 3674 & 2.99$\pm0.82$ &  -0.1$\pm$0.08  &  -0.35$\pm$0.15  & -0.13$\pm$0.09 & -0.14$\pm$0.12 & -0.07$\pm$0.12 & 0.00$\pm$0.03 \\
All AGN    & 298  & 2.99$\pm0.63$ &  \textbf{0.94$\pm$0.28}  &  \textbf{1.53$\pm$0.50}  & \textbf{2.10$\pm$0.32} & \textbf{1.39$\pm$0.44} & 1.29$\pm$0.45 & \textbf{0.43$\pm$0.11} \\
X-ray AGN  & 240  & 2.99$\pm0.64$ &  \textbf{0.93$\pm$0.31}  &  1.65$\pm$0.57  & \textbf{1.92$\pm$0.35} & \textbf{1.37$\pm$0.46} & 1.25$\pm$0.48 & \textbf{0.37$\pm$0.12} \\
Radio AGN  & 108  & 2.99$\pm0.58$ &  \textbf{1.55$\pm$0.44}  &  \textbf{2.71$\pm$0.85}  & \textbf{2.67$\pm$0.54} & \textbf{2.56$\pm$0.68} & 1.94$\pm$0.71 & \textbf{0.56$\pm$0.18} \\

\hline
\multicolumn{9}{l}{\textit{Ly}$\alpha$ Luminosity (erg s$^{-1}$)} \\
\hline

42.0 $<$ $\log_{10}(L_{\rm Ly\alpha}) \leq$ 43.0 & 2640 & 2.99$\pm0.65$ & -0.18$\pm$0.09 & -0.44$\pm$0.18 & -0.24$\pm$0.11 & -0.35$\pm$0.14 & -0.22$\pm$0.15 & -0.03$\pm$0.04 \\
43.0 $<$ $\log_{10}(L_{\rm Ly\alpha}) \leq$ 43.3 & 748  & 3.33$\pm0.89$ & -0.02$\pm$0.17 & -0.19$\pm$0.33 & -0.08$\pm$0.21 & -0.06$\pm$0.26 & 0.02$\pm$0.28 & 0.07$\pm$0.07 \\
43.3 $<$ $\log_{10}(L_{\rm Ly\alpha}) \leq$ 43.5 & 170  & 4.12$\pm1.10$ & 0.26$\pm$0.30 & -0.09$\pm$0.69 & 0.18$\pm$0.44 & 0.20$\pm$0.57 & 0.79$\pm$0.59 & 0.05$\pm$0.15 \\
43.5 $<$ $\log_{10}(L_{\rm Ly\alpha}) \leq$ 43.8 & 83   & 3.33$\pm1.11$ & 0.83$\pm$0.50 & 2.22$\pm$0.94 & 0.93$\pm$0.62 & 1.41$\pm$0.78 & 1.84$\pm$0.82 & 0.46$\pm$0.21 \\
43.5 $<$ $\log_{10}(L_{\rm Ly\alpha}) \leq$ 43.8 excl.\ AGN$^b$ & 50 & 4.58$\pm1.07$ & -0.14$\pm$0.63 & 1.23$\pm$1.24 & -0.52$\pm$0.79 & 0.88$\pm$1.03 & 1.46$\pm$1.05 & 0.42$\pm$0.27 \\
43.8 $<$ $\log_{10}(L_{\rm Ly\alpha}) \leq$ 44.8 & 33 & 3.12$\pm0.75$ & 0.75$\pm$0.80 & 1.42$\pm$1.54 & 1.02$\pm$1.01 & 0.09$\pm$1.27 & 1.59$\pm$1.28 & 0.46$\pm$0.31 \\

\hline
\multicolumn{9}{l}{Stellar Mass ($M_\star$)} \\
\hline

7.8 $<$ $\log_{10}(M_\star) \leq$ 9.0 & 1047 & 2.99$\pm0.52$ & -0.34$\pm$0.15 & -0.61$\pm$0.28 & -0.58$\pm$0.17 & -0.58$\pm$0.21 & -0.43$\pm$0.23 & -0.09$\pm$0.06 \\
9.0 $<$ $\log_{10}(M_\star) \leq$ 9.5 & 1115 & 2.99$\pm0.78$ & -0.36$\pm$0.14 & -1.14$\pm$0.27 & -0.65$\pm$0.17 & -0.57$\pm$0.22 & -0.39$\pm$0.23 & 0.02$\pm$0.06 \\
9.5 $<$ $\log_{10}(M_\star) \leq$ 10.0 & 843  & 3.16$\pm0.93$ & -0.06$\pm$0.16 & -0.23$\pm$0.32 & 0.05$\pm$0.20 & -0.28$\pm$0.25 & -0.20$\pm$0.26 & -0.04$\pm$0.06 \\
10.0 $<$ $\log_{10}(M_\star) \leq$ 12.0 & 669 & 3.16$\pm0.97$ & \textbf{0.70$\pm$0.19} & \textbf{1.22$\pm$0.36} & \textbf{1.85$\pm$0.22} & \textbf{1.82$\pm$0.30} & \textbf{1.66$\pm$0.30} & \textbf{0.29$\pm$0.07} \\
10.0 $<$ $\log_{10}(M_\star) \leq$ 12.0 excl.\ AGN$^b$ & 461 & 3.33$\pm1.05$ & 0.60$\pm$0.21 & 0.77$\pm$0.43 & \textbf{1.27$\pm$0.28} & \textbf{1.80$\pm$0.35} & \textbf{1.82$\pm$0.36} & 0.16$\pm$0.09 \\

\hline
\multicolumn{9}{l}{Redshift ($z$)} \\
\hline

2.2 $<$ $z \leq$ 2.7 & 876  & 2.50$\pm0.10$ & 0.00$\pm$0.15 & -0.35$\pm$0.29 & 0.03$\pm$0.19 & -0.21$\pm$0.25 & 0.07$\pm$0.26 & -0.04$\pm$0.06 \\
2.7 $<$ $z \leq$ 3.5 & 2139 & 3.16$\pm0.18$ & -0.09$\pm$0.10 & -0.17$\pm$0.18 & -0.13$\pm$0.12 & -0.22$\pm$0.16 & -0.19$\pm$0.16 & -0.01$\pm$0.04 \\
3.5 $<$ $z \leq$ 6.0 & 659  & 4.82$\pm0.67$ & -0.25$\pm$0.16 & -0.77$\pm$0.31 & -0.33$\pm$0.19 & -0.01$\pm$0.26 & 0.15$\pm$0.28 & 0.10$\pm$0.07 \\

\hline
\end{tabular}

\vspace{2mm}

$^a$ Number of LAEs included in each stack.
$^b$ LAEs containing radio and/or X-ray AGN are excluded from these stacks.
$^c$ Median redshift of the stacks; uncertainties represent the standard deviation (i.e. the spread) of the redshift distribution of the stacked sample.
\end{table*}

\begin{figure*}
\begin{center}
\includegraphics[scale=0.5]{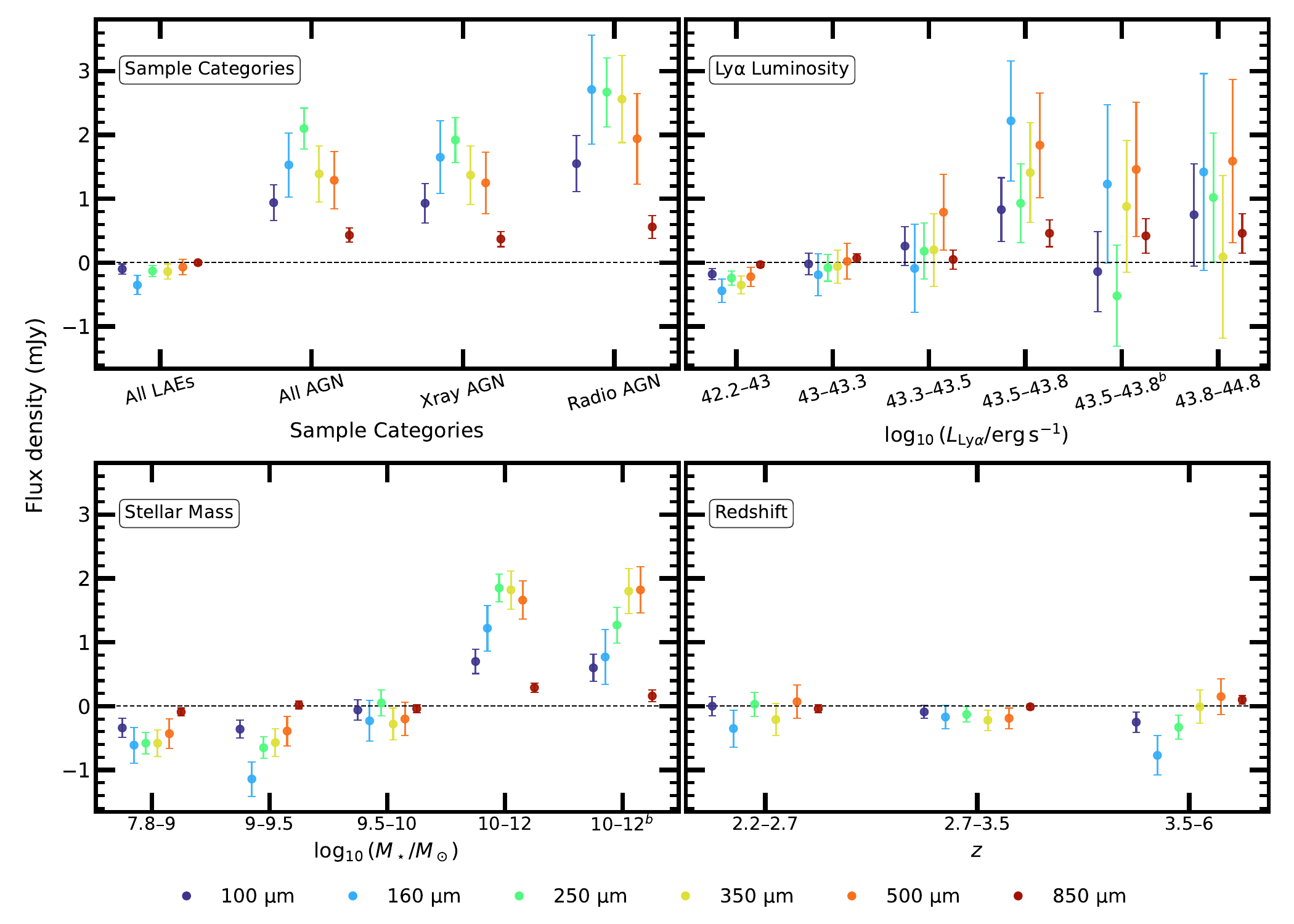}
\caption{The far-infrared flux densities at 100, 160, 250, 350, 500 and 850\,\mic, measured from our median stacks of all SC4K LAEs and different subsamples. They are shown as purple, light blue, light green, yellow, orange and dark red circles, respectively. Each panel presents results from stacks of different sample categories (\textit{top left}), and subsamples binned by \lya\ luminosity (\textit{top right}), stellar mass (\textit{bottom left}), and redshift (\textit{bottom right}). $1\sigma$ uncertainties are indicated by the error bars, which demonstrates that at all wavelengths the stacks of all LAEs, and the samples split by \lya\ luminosity and redshift are undetected at the $3\sigma$ level. However $\geq 3\sigma$ signals are observed in the stacks of all AGN, X-ray AGN, radio AGN and high stellar mass ($\mstar = 10^{10\--12}$ \msun) populations at  most wavelengths. Even when AGN are excluded stacks of the most massive LAEs (i.e.\ $\mstar = 10^{10\--12}$\,\msun)  have a 2.9$\sigma$ signal at 100 \mic\ and are detected at $\geq 4\sigma$ signals at 250, 350 and 500 \mic. 
A tentative trend of higher average flux densities for LAEs with \lya\ luminosities $\gtrsim 10^{43.5}\,\mathrm{erg\,s^{-1}}$ is observed (though with higher noise), including 2.4, 2.2 and $2.2\sigma$ signals in the 160, 500 and $850\,\mu\mathrm{m}$ stacks for $L_{\mathrm{Ly}\alpha} = 10^{43.5}$--$10^{43.8}\,\mathrm{erg\,s^{-1}}$ (Figure~\ref{fig:stacks}). 
All stacks exclude the five individually detected and four blended LAEs (Section~\ref{sec:indivduals}). The superscript $^b$ indicates subsets from which AGN are excluded. }
\label{fig:stack_depth}
\end{center}
\end{figure*}

\subsubsection{Stacking Results}
We stack the whole population of LAEs, and also separately stack LAEs with radio and/or X-ray AGN, as well as subsets that are binned by \lya\ luminosity, stellar mass and redshift. The binning scheme from \citet{Calhau2020} is adopted, such that our results can be directly compared with the \citet{Calhau2020} X-ray stacks. The subsets are listed in Table~\ref{tab:Stack_flux} along with their median stacked flux densities in the 100--850\,\mic\ data. Figure~\ref{fig:stack_depth} displays these measurements graphically and Figure~\ref{fig:stacks} shows example stacks including all LAEs, LAEs that contain AGN, LAEs with $\log_{10}({\rm L_{Ly\alpha}/ erg\,s}^{-1})=43.5$--43.8, those with  stellar mass  $\log_{10}({\rm M_{\star}/M_{\odot})=10}$--12  (both including and excluding AGN) and those at $z=3.5$--6. 

We do not detect far-infrared emission from the population of all 3674 LAEs as a whole, suggesting that their median flux density is  $<0.24$\,mJy, $<0.27$\,mJy, and $<0.09$\,mJy ($3\sigma$) at 100, 250, and 850\,\mic, respectively. Most of the other stacks are also undetected at all targeted wavelengths, and upper limits are used for these throughout our analyses (Section~\ref{Sec:4}). However, the LAEs that contain AGN or are the most massive are detected at most of the targeted far-infrared wavelengths i.e.\ these are the LAEs that contain the most dust, as shown in Figure~\ref{fig:stack_depth} and Table~\ref{tab:Stack_flux}. 
The second brightest bin of \lya\ luminosity (i.e.\ with $\log_{10}({\rm L_{Ly\alpha}/ erg\,s}^{-1})=43.5$--43.8) is undetected at 100, 250 and 350\,\mic, but has tentative signals at the $2.4$, $2.2$ and $2.2\sigma$ levels at  160, 500 and 850\,\mic, respectively. This bin contains $\sim2.5\times$ more LAEs than the highest luminosity bin and therefore reaches $\sim1.6\times$ deeper, so these results are consistent with \lya\ luminosity also being a driver of far-infrared emission in LAEs. 

Since AGN (classified by X-ray and/or radio detections) are more common in the most massive and most \lya-luminous LAEs \citep{Calhau2020} we next exclude AGN and repeat the stacking procedure for the \lya\ luminosity, stellar mass and redshift bins. As expected, subsets that were not previously detected remain undetected when AGN are excluded. The 50 LAEs with $\log_{10}({\rm L_{Ly\alpha}/ erg\,s}^{-1})=43.5$\,--\,43.8 that are not AGN are not detected when stacked, even when combined with the brightest non-AGN LAEs, which increases the sample to 59 and improves the sensitivity. However, the stack of the most massive LAEs is detected at 4.5, 5.1 and $5.1\sigma$ at 250, 350 and 500\,\mic, respectively, and has a tentative $2.9\sigma$ signal at 100\,\mic\ even after removing AGN. This suggests that both AGN and stellar mass drive the far-infrared emission from LAEs. 

\begin{figure*}
\begin{center}
\includegraphics[scale=0.71]{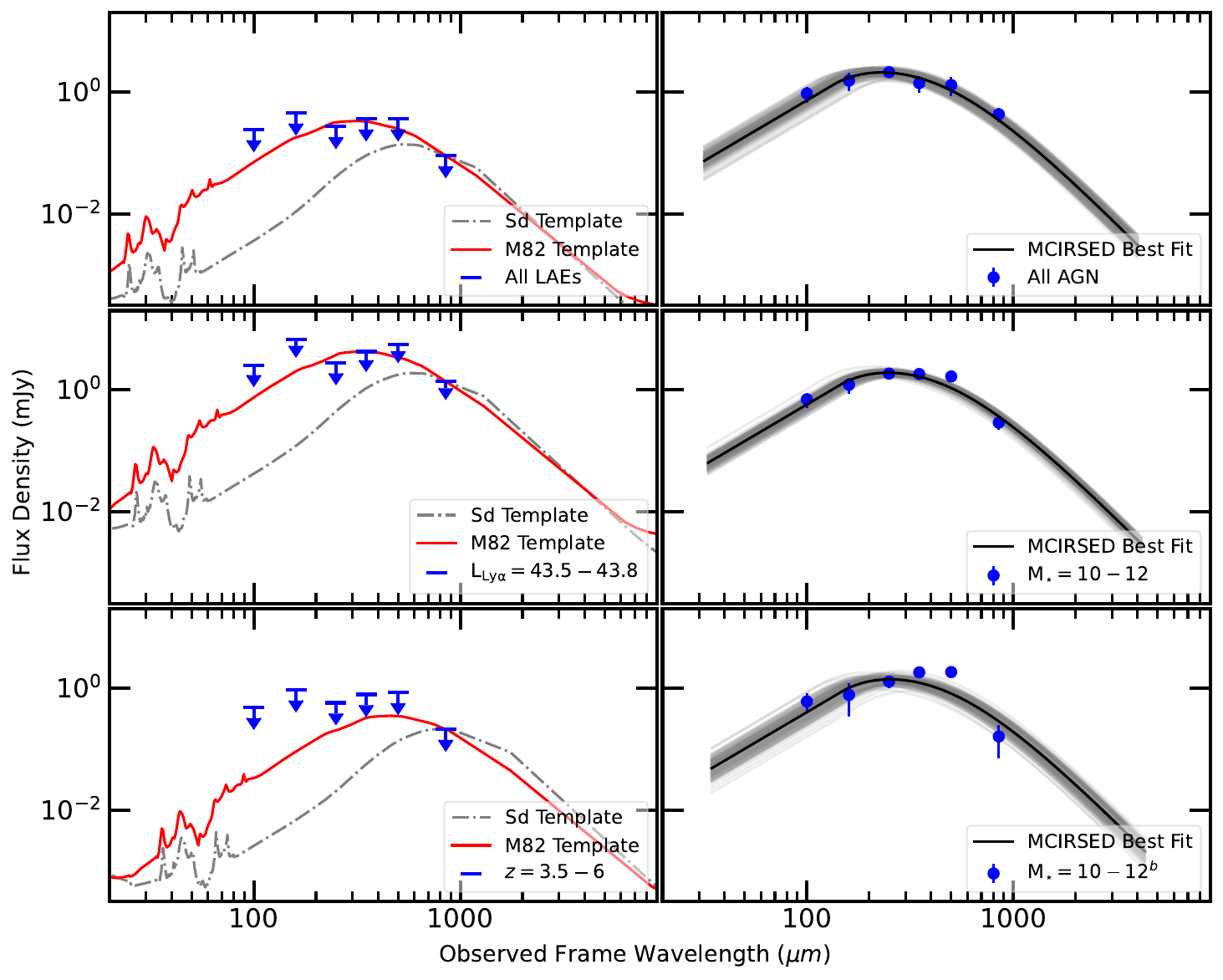}
\caption{Far-infrared SEDs for key stacks from Table~\ref{tab:Stack_flux}. The left-hand column shows stacks that are undetected at all bands, overlaid with the SWIRE Sd and M82 templates \citep{Polletta2007} that provide upper limits on the infrared luminosities. The right-hand column shows stacks that have $\ge2$ detections and are fit with MCIRSED as described in Section~\ref{sec:sedfit}. For these SEDs the black line represents the best fit, while the gray lines show 100 other less probable fits sampled from the MCMC posterior distribution. Upper limits are plotted at the $3\sigma$ level. The left-hand panels show, from top to bottom, stacks of all SC4K LAEs, those with ${\rm L_{Ly\alpha}} = 10^{43.5\--43.8}$\,erg s$^{-1}$, and those at  $z=3.5\--6$. In the right-hand column the stacks of all LAEs containing radio and/or X-ray AGN, the most massive LAEs ($\mstar = 10^{10\--12}$ \msun) and the most massive LAEs excluding those that contain AGN (denoted by $^b$). The SEDs demonstrate the range of infrared luminosities probed by the Sd and M82 templates, the robustness of the stacked flux densities and the quality of the MCIRSED fits.} 
\label{fig:SED_stacks}
\end{center}
\end{figure*}

\section{Results And Discussion} 
\label{Sec:4}

\subsection{SED Fitting} 
\label{sec:sedfit}

Having identified far-infrared emission from some individual LAEs and measured median flux densities from LAE stacks, we next use SED fitting to quantify the dust emission from these galaxies. For the five individually-detected LAEs the photometry used is presented in Table~\ref{tab:six_LAEs_flux1} and for stacks we use the detections and limits given in Table~\ref{tab:Stack_flux}. Of the 3683 sources studied here, 359 LAEs have narrow-band-constrained photometric redshifts, while 3324 LAEs have medium-band-constrained photometric redshifts \citep{Sobral2018}. We employ spectroscopic redshifts for two far-infrared detected LAEs (Table~\ref{tab:six_LAEs_flux1}) reported in \citet{Hasinger2018} and \citet{DESI_Collaboration2024}. For stacked samples the median redshifts of the included sources are used (Table~\ref{tab:Stack_flux}). 

Different methods of SED fitting are used depending on whether or not the system is detected at  $\geq 3\sigma$ in $\geq 2$ wavelengths. For those stacks or sources that are undetected or only detected in one band we use the SWIRE Sd and M82 galaxy SED templates \citep{Polletta2007} to constrain the far-infrared luminosity. These SED templates are scaled to the $3\sigma$ flux density upper limit at 850\,\mic\ (which, relative to the SEDs, is the deepest of the datasets; Figure~\ref{fig:SED_stacks}) at the redshift of the individual LAE or the median redshift value of the stack. The Sd and M82 templates are chosen because they cover the full range of effective dust temperatures in SWIRE ($T_{{\rm dust}}\sim$\,20--60K), and therefore encompass the range of plausible infrared luminosities. The temperature range covered by these templates is consistent with high-redshift studies \citep[e.g.,][]{Magdis2012, Faisst2020, Sommovigo2022} and with simulation results from \citet{Liang2019}. For systems that have multiple $\ge3\sigma$ detections the SED fitting code {\sc Monte Carlo Infrared Spectral Energy Distribution} (MCIRSED; \citealt{Casey2012, Drew2022}) is used. There are six parameters in the MCIRSED: the greybody normalization ($N_{\mathrm{bb}}$), power-law normalization ($N_{\mathrm{pl}}$), greybody temperature ($T$), emissivity index ($\beta_{\rm IR}$), mid-infrared power-law slope ($\alpha$), and mid-infrared turnover wavelength ($\lambda_c$). Since $N_{\mathrm{pl}}$ and $\lambda_c$ are tied to fitting values of $\alpha$, $N_{\mathrm{bb}}$ and $T$, there are effectively four free-parameters. Since we have only 2 to 3 photometric points with signal to noise ratios $\geq3\sigma$ in most cases, we reduce the number of degrees of freedom by fixing the emissivity index to  $\beta_{\rm IR}= 1.5$ as suggested by \citet{Casey2012} and commonly adopted in the literature \citep[e.g.,][]{Chapman2005, Pope2006, Casey2009} and setting the mid-infrared power-law slope to $\alpha = 2.0$ \citep{Casey2012}. We treat $T$ and $N_{\mathrm{bb}}$ as free parameters. We also tested the effect of not fixing $\alpha$ and $\beta_{\rm IR}$ and find no significant differences in the derived far-infrared luminosities, but significantly increased uncertainty on this parameter. 

For the {\sc MCIRSED} fits (i.e.\ individual LAE and stacks with $\ge2$ bands  with $\ge3\sigma$ detections) the SED fitting is performed using the measurements in all detected bands and we include the measured values (and uncertainties) for any remaining band(s) that have $\geq 2.5\sigma$ signals. For any values that are below the $2.5\sigma$ level we use the $2.5\sigma$ noise as an upper limit. For the LAE SC4K-IA574-34828 reliable flux density measurement at 350\,\mic\ is not available \citep{Jin2018} and for SC4K-IA484-69327 the catalogued $1\pm5$\,mJy flux density at 250\,\mic\ is not inconsistent with other photometry for this source (Table~\ref{tab:six_LAEs_flux1}) and we therefore exclude these data from the SED fits. {\sc MCIRSED} and SWIRE template-based SEDs for key stacks are presented in Figure~\ref{fig:SED_stacks}. From the SED fits described above we estimate the rest-frame 8--1000\,\mic\ far-infrared luminosities ($\lfir $), including $3\sigma$ upper limits for LAE stacks that are not detected in \herschel\ or SCUBA-2. The SED-derived far-infrared properties of the five individually-detected LAEs are presented in Table~\ref{tab:six_LAEs_flux2}.  Table~\ref{tab:stack_properties} presents results for the stacked data. 

We next derive dust-obscured SFRs (${\rm SFR_{FIR}}$) for the LAEs and stacks using the relationship between infrared luminosity and SFR from \citet{Kennicutt1998} which is given in Equation~\ref{eq:lirsfr}:
\begin{equation} \label{eq:lirsfr}
\text{SFR}_{\text{FIR}} (\rm{M}_{\sun} \; yr^{-1}) =  1.09 \times 10^{-10}  L_{FIR}(\rm{L}_{\sun})
\end{equation}

Equation~\ref{eq:lirsfr} assumes \citet{Salpeter1955} IMF, so we multiply the resulting SFRs by a factor of 0.63 to scale them to a \citet{Chabrier2003} IMF. For systems that are not detected in the infrared Equation~\ref{eq:lirsfr} is used to calculate the $3\sigma$ upper limit on the obscured SFR. 

The far-infrared luminosities of individually-detected LAEs are all in the ULIRG range with $\lfir=10^{12.5}$ to $10^{13.1}$\,\lsun, corresponding to obscured SFRs of 360--1250\,\myr\ (assuming no AGN contribution to the infrared fluxes). 
The average far-infrared luminosities of the subsets of LAEs that are detected in the stacks are slightly lower, in the range $10^{11.9}$ to $10^{12.3}$\,\lsun, corresponding to far-infrared-based SFRs of  100--190\,\myr. 
LAEs that are in stacks that are not detected are fainter and have $3\sigma$ upper limits of $\lfir <10^{10.4}$ to $10^{12.3}$\,\lsun, equivalent to  SFR$_{\rm{FIR}}= 3$--230\,\myr. In previous studies, \citet{Wardlow2014} stacked $z=2.1$--4.5 LAEs at SPIRE and SCUBA-2 wavelengths, and \citet{Haruka2015} stacked $z=2.23$ LAEs at PACS and MIPS wavelengths. These studies constrained the average emission of LAEs at these redshifts to $\lfir <10^{10.04}$\,\lsun\ and $\lfir \lesssim10^{11}$\,\lsun\ (depending on the redshift and assumed SED) for \citet{Haruka2015} and \citet{Wardlow2014}, respectively. Our measurements are consistent with these results, giving e.g.\ $\lfir\lesssim\times10^{11.4}$\,\lsun\ for $z=2.2$--2.7 LAEs scaled to the M82 template and $\lfir \lesssim\times10^{11.1}$\,\lsun\  for all SC4K LAEs and the same template. These results demonstrate that whilst most LAEs are faint in the infrared, with obscured SFRs at the level of 10s of \myr\ or lower. Nevertheless, some LAEs can be infrared-luminous, and these are typically those that are massive or that contain AGN. For the stack of all AGN LAEs (Table~\ref{tab:stack_properties}), we estimate the potential impact of AGN contamination on SFR$_{\rm{FIR}}$ by fitting SWIRE AGN templates \citep{Polletta2007} to the observed far-infrared photometry, in addition to the MCIRSED star-forming SED. We use the unobscured QSO1 and obscured QSO2 templates from SWIRE to estimate the range of possible AGN contribution. Each AGN template is redshifted and scaled to the observed photometry to derive its infrared luminosity contribution. From this, we estimate that the AGN can contribute $\sim$50\% and $\sim$40\% of the total infrared emission used to derive SFR$_{\rm{FIR}}$ in the cases of QSO1 and QSO2, respectively.

\begin{table*}
\caption{Physical properties of five individually detected LAEs.}
\label{tab:six_LAEs_flux2}
\begin{tabular}{lcccccc} 
\hline
Source ID & $\log_{10}$($M_{\star}$)$^h$  &  $\log_{10}$(\lfir)$^c$   & SFR$_{\rm{FIR}}$$^c$ & SFR$_{\rm{UV}}$$^d$ & SFR$_{\rm{Ly\alpha}}$$^d$ & \fesc$^d$  \\
 & (\msun) & ($\mathrm{L_\odot}$) & ($\mathrm{M_\odot\,yr^{-1}}$) & ($\mathrm{M_\odot\,yr^{-1}}$) & ($\mathrm{M_\odot\,yr^{-1}}$) & \\
\hline
SC4K-IA505-178627 & 10.9±0.1  & $12.5\pm0.1$ & $360^{+40}_{-50}$    & 35±4   & 28±1 & 0.070±0.009 \\
SC4K-IA574-34828  & 11.1±0.1       & $13.0\pm0.1$ & $1090^{+220}_{-220}$ & 10±1   & 7±1  & 0.006±0.001 \\
SC4K-IA484-28746  & 11.5±0.1       & $12.9\pm0.1$ & $970^{+120}_{-280}$  & 11±1   & 14±1 & 0.014±0.002 \\
SC4K-IA484-69327  & 10.8±0.1     & $13.0\pm0.1$ & $1110^{+260}_{-250}$ & 106±12 & 53±1 & 0.043±0.009 \\
SC4K-IA679-223923 & 10.6±0.1     & $13.1\pm0.1$ & $1250^{+220}_{-200}$ & 7±1    & 11±2 & 0.009±0.002 \\
\hline
\end{tabular}
\\
$^c$ IR luminosities ($\lfir $) and SFRs (${\rm SFR_{FIR}}$) are derived in Section~\ref{sec:sedfit}.
$^d$ UV and \lya-derived SFRs and the calculation of \fesc\ are described in Section~\ref{sec:fesc}.
$^h$ Stellar mass from \citet{Santos2020}, mentioned in Section~\ref{Sec:LAE_sample}.
\end{table*}

\begin{table*}
\centering
\caption{Average far-infrared properties of different groups of LAEs derived from SED fitting to our stacked data. Limits are at the $3\sigma$ level and the two values are derived using the Sd (first value) and M82 (second value) SED templates. 
}
\label{tab:stack_properties}
\begin{tabular}{lcccccccr} 
\hline 
Sample                    & \# Sources$^a$  & $z_{\rm median}$$^c$  &  $\log_{10}$(\lfir)$^d$  &   SFR$_{\rm{FIR}}$$^e$   &   SFR$_{\rm{UV}}$$^f$      &   SFR$_{\rm{Ly\alpha}}$$^g$  & \fesc$^h$   \\
	                  &             &         &      (\lsun)          &  (\myr)         &  (\myr)           &  (\myr)             &         \\
\hline 
All LAEs             & 3674           &  2.99$\pm0.82$   & $\leq$10.4, $\leq$11.1          & $\leq$3, $\leq$15   & 4±1 & $4.1\pm0.6$ & $\geq$0.59, $\geq$0.21 \\ 
All AGN                   & 298          &  2.99$\pm0.63$  & $12.1\pm0.1$ & $120\pm20$ & 6±1  & $6.4\pm0.6$ & $0.05\pm0.01$  \\ 
X-ray AGN                 & 240          &  2.99$\pm0.64$   & $12.1\pm0.1$ & $120\pm20$ & 7±01  & $7.1\pm0.6$ & $0.05\pm0.01$ \\ 
Radio AGN                 & 108          &   2.99$\pm0.58$     & $12.3\pm0.1$ & $190\pm40$ & 6±1  & $7.1\pm0.6$ & $0.04\pm0.01$ \\ 
\hline
\multicolumn{0}{c}{\lya\ Luminosity (erg s$^{-1}$)} \\ 
\hline
42.0 < $\log_{10}(\rm{L}_{\rm{Ly\alpha}})$  $\leq$  43.0   & 2640  & 2.99$\pm0.65$ & $\leq$10.5, $\leq$11.3 & $\leq$ 4, $\leq$20 & 3±1 & $3.5\pm0.6$  & $\geq$0.48, $\geq$0.15  \\ 
43.0 < $\log_{10}(\rm{L}_{\rm{Ly\alpha}})$  $\leq$  43.3   & 748   & 3.33$\pm0.89$ & $\leq$10.8, $\leq$11.5 & $\leq$7, $\leq$40 & 7±2   & $7.6\pm0.7$ & $\geq$0.55, $\geq$0.18  \\ 
43.3 < $\log_{10}(\rm{L}_{\rm{Ly\alpha}})$  $\leq$  43.5   & 170   & 4.12$\pm1.10$ & $\leq$11.2, $\leq$11.8 & $\leq$20, $\leq$70 & 10±2  & $13.4\pm1.0$ & $\geq$0.48, $\geq$0.16  \\ 
43.5 < $\log_{10}(\rm{L}_{Ly\alpha})$  $\leq$  43.8 & 83    & 3.33$\pm1.11$ & $\leq$11.6, $\leq$12.3 & $\leq$50, $\leq$230 & 17±4  & $22.8\pm0.5$ & $\geq$0.36, $\geq$0.09 \\ 
43.5 < $\log_{10}(\rm{L}_{Ly\alpha})$  $\leq$  43.8 excl.\ AGN$^b$ & 50  & 4.58$\pm1.07$ & $\leq$11.5, $\leq$12.1 & $\leq$30, $\leq$130 & 15±3  & $21.8\pm1.6$ & $\geq$0.45, $\geq$0.15 \\ 
43.8 < $\log_{10}(\rm{L}_{Ly\alpha})$  $\leq$  44.8   & 33    & 3.12$\pm0.75$ & $\leq$11.4, $\leq$12.1 & $\leq$30, $\leq$160 & 40±10   & $51.1\pm1.2$ & $\geq$0.76, $\geq$ 0.27   \\ 
\hline
\multicolumn{0}{c}{Stellar Mass (\msun)} \\ 
\hline
7.8 < $\log_{10}(\rm{M_{\star}})$  $\leq$  9.0        & 1047   & 2.99$\pm0.52$ & $\leq$10.7, $\leq$11.4 & $\leq$10, $\leq$30 & 3±1 & $3.4\pm0.6$ &  $\geq$0.42, $\geq$0.11 \\ 
9.0 < $\log_{10}(\rm{M_{\star}})$  $\leq$  9.5        & 1115   & 2.99$\pm0.78$ & $\leq$10.7, $\leq$11.4 & $\leq$10, $\leq$30  & 5±1 & $4.2\pm0.6$ & $\geq$0.41, $\geq$0.12 \\
9.5 < $\log_{10}(\rm{M_{\star}})$  $\leq$  10.0       & 843    & 3.16$\pm0.93$ & $\leq$10.7, $\leq$11.4 & $\leq$10, $\leq$30 & 10±1 & $4.1\pm0.6$ & $\geq$0.37, $\geq$0.12 \\ 
10.0< $\log_{10}(\rm{M_{\star}})$  $\leq$  12.0   & 669    & 3.16$\pm0.97$ & $12.0\pm0.1$ & $120\pm15$ & 7±1 & $5.7\pm0.7$ & $0.05\pm0.01$ \\  
10.0< $\log_{10}(\rm{M_{\star}})$  $\leq$  12.0 excl.\ AGN$^b$ & 461  & 3.33$\pm1.05$ & $11.9\pm0.1$ & $100\pm20$ & 7±1 & $5.0\pm0.6$ & $0.05\pm0.01$ \\ 
\hline
\multicolumn{0}{c}{Redshift (\textit{z}) } \\ 
\hline
2.2 < \textit{z}  $\leq$  2.7   & 876    & 2.50$\pm0.10$ & $\leq$10.7, $\leq$11.4 & $\leq$5, $\leq$30 & 3±1 & $2.5\pm0.4$ & $\geq$0.33, $\geq$0.08 \\ 
2.7 < \textit{z}  $\leq$  3.5   & 2139   & 3.16$\pm0.18$ & $\leq$10.6, $\leq$11.3 & $\leq$5, $\leq$20 & 4±1 & $4.2\pm0.6$ & $\geq$0.54, $\geq$0.18  \\ 
3.5 < \textit{z}  $\leq$  6.0   & 659    & 4.82$\pm0.67$ & $\leq$10.9, $\leq$11.5 & $\leq$10, $\leq$30 & 9±2 & $7.4\pm1.1$ & $\geq$0.40, $\geq$0.17 \\ 
\hline 
\end{tabular}

$^a$ Number of LAEs included in each stack.
$^b$ LAEs containing radio and/or X-ray AGN are excluded from these stacks. 
$^c$ Median redshift of the stacks; uncertainties represent the standard deviation (i.e. the spread) of the redshift distribution of the stacked sample.
$^d$ Rest-frame 8--1000\,\mic\ far-infrared luminosity. 
$^e$ Dust-obscured SFR, derived from the far-infrared luminosity (Section~\ref{sec:sedfit}). 
$^f$ Unobscured SFR, derived from dust-uncorrected rest-frame UV luminosity (Section~\ref{sec:fesc}). 
$^g$ Apparent \lya-derived SFR, calculated from the observed \lya\ luminosity (Section~\ref{sec:fesc}). 
$^h$ \lya\ escape fraction, calculated as described in Section~\ref{sec:fesc}.
\end{table*}
\begin{figure}
\begin{center}
\includegraphics[scale=0.375]{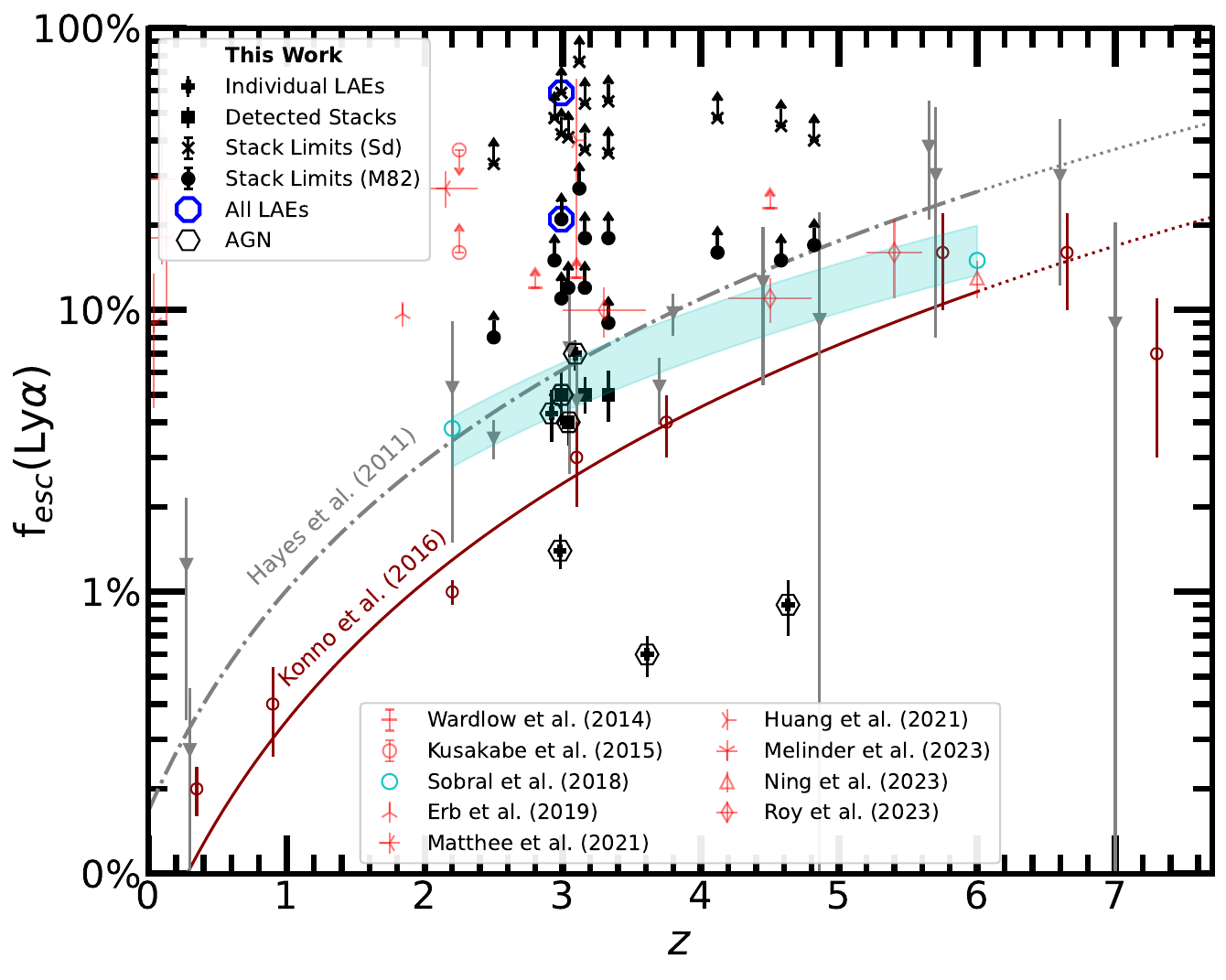}
\caption{The evolution of \lya\ escape fraction as a function of redshift, with our results shown in black. We highlight the results from stacking the whole SC4K catalogue and indicate LAEs containing AGN and the AGN-specific stacks. For stacks that are not detected in the infrared data we show the $3\sigma$ lower limits derived assuming both the Sd and M82 templates, where the Sd templates give higher \fesc\ limits for the same dataset. The grey symbols and line shows results and the fit from \citet{Hayes2011}, the maroon symbols and line are the equivalent from \citet{Konno2016}; dotted portions of the lines indicate our extrapolation of these fits to higher redshifts than originally studied.
Red symbols represent \fesc\ from \citet{Wardlow2014, Haruka2015, Erb2019, Huang2021, Matthee2021, Melinder2023, Ning2023} and \citet{Roy2023}. The cyan circles and cyan-shaded region show the range of \fesc\ for the entire SC4K LAE sample \citep{Sobral2018}, inferred from the ratio of observed \lya\ luminosity density to intrinsic \lya\ production estimated from UV SFR density. Directly overlapping points are shifted to left or right by $\Delta z$=0.05 to aid visibility. \fesc\ range from $\fesc=1$--7\% for detections (both individual and stacked), while upper limits exceed $21\%$, with the overall scatter suggesting a clumpy dust distribution in the ISM of LAEs. However, many of our stacks probed deeper than previous studies of \fesc\ and still remain undetected, which could indicate significant evolution or variety in the population and suggests that deeper studies of individual sources are required. 
}
\label{fig:escape_fraction}  
\end{center}
\end{figure}

\subsection{\lya\ Escape Fraction} \label{sec:fesc}
\lya\ is affected by dust attenuation and resonant scattering by neutral hydrogen, which cause the absorption and re-emission of  \lya\ photons at different wavelengths and in different directions. These effects make \lya\ observations hard to interpret, particularly for measuring SFR, for which \lya\ is often used at high redshift \citep[e.g.][]{Tapken2007, Floranes2014, Oyarz2017}. A key parameter is \fesc, which measures the fraction of  \lya\ produced  that is detected. Thus, \fesc\ can be used to calculate the true \lya\ emission, accounting for attenuation and scattering. Note that it is possible that the regions from which we can detect \lya\ may be completely disparate from those that emit detectable infrared radiation: this is particularly probable at high-redshift, where star formation is known to be ``clumpy'' \citep[e.g.][]{Dekel2009}. Therefore, to estimate the total SFR of a galaxy (and thus the \lya\ production) requires the sum of both the obscured and unobscured SFR. \fesc\ can then be calculated by comparing the total SFR with the SFR inferred from the \lya\ luminosity as in Equation~\ref{eq:fesc}:
\begin{equation} \label{eq:fesc}
\text{\fesc} = \frac{\text{SFR}_{\text{Ly}\alpha}}{\text{SFR}_\text{total}}
= \frac{\text{SFR}_{\text{Ly}\alpha}}{\text{SFR}_{\text{UV}} + \text{SFR}_{\text{FIR}}}
\end{equation} 
where $\text{SFR}_{\text{total}}$ is defined as the sum of the unobscured ($\text{SFR}_{\text{UV}}$) and the obscured ($\text{SFR}_{\text{FIR}}$). $\text{SFR}_{\text{FIR}}$ is calculated from our infrared observations as described in Section~\ref{sec:sedfit}. For the UV-based SFR we use the rest-frame $\lambda_{\rm UV, 0}=1400$--1600\,{\AA} UV absolute magnitude (M$_{\rm{UV}}$) from \citet{Santos2020}, which is extracted from {\sc magphys} fitting with a prior $\beta_{\rm UV} \ge -2.44$ and has not been dust-corrected. The observed \lya\ luminosity (without dust correction) is also taken from \citet{Santos2020}. SFRs derived from both the \lya\ luminosity (${\rm L_{Ly\alpha}}$) and ${\rm M_{UV}}$ are calculated according to \citet{Kennicutt1998}, as given in Equations~\ref{eq1} and \ref{eq2}:
\begin{equation} \label{eq1}
\text{SFR}_{\text{Ly}\alpha} (\rm{M}_{\sun} \; yr^{-1}) = 5.73 \times 10^{-43} L_{\text{Ly}\alpha}(\text{erg}s^{-1})
\end{equation}
\begin{equation} \label{eq2}
\text{SFR}_{\text{UV}} (\rm{M}_{\sun} \; yr^{-1}) = 5.4 \times 10^{-8} \times 10^{-0.4 \text{M}_{\rm{UV}}}
\end{equation}
As in Equation~\ref{eq:lirsfr} these SFRs are based on a \citet{Salpeter1955} IMF, so we multiply the resulting SFRs by a factor of 0.63 to scale them to a \citet{Chabrier2003} IMF. SFRs calculated from the different tracers and the resulting \fesc\ are presented in Table~\ref{tab:six_LAEs_flux2} for the LAEs that are individually-detected in the infrared, and in Table~\ref{tab:stack_properties} for the stacked data. 

Figure~\ref{fig:escape_fraction} shows the redshift evolution of \fesc\ for SC4K LAEs as calculated here from single-dish submillimetre data, compared with previous studies \citep{Wardlow2014, Haruka2015, Erb2019, Matthee2021, Huang2021, Melinder2023, Ning2023, Roy2023},  and parametrisations of the global trend  \citep{Hayes2011, Konno2016}. The \citet{Hayes2011} and \citet{Konno2016} are offset from each other due to their uses of different \lya\ and UV luminosity limits in the calculation of the  \lya\ and UV SFR density from \lya\ and UV luminosity functions. We also compare our results with the \fesc\ for the full SC4K sample \citep{Sobral2018} derived from the ratio of observed \lya\ luminosity density to the intrinsic \lya\ production inferred from UV SFR density and global ionization efficiency (case-B recombination).

The five LAEs that are individually-detected in the submillimetre data all have \fesc\ from 1 to 7\%, values which are scattered on or below the global \fesc\ and \citet{Sobral2018} at their redshifts of $z\sim3$\,--\,4.6 (Figure~\ref{fig:escape_fraction}). However, we caution that all five of these LAEs contain AGN and we have assumed no AGN contribution to their far-infrared emission. A notable tension with this assumption arises from the high dust temperature inferred from the MCIRSED fits (Figure~\ref{fig:SED_Individual}). The median dust temperature of the the five sources is $\sim$70\,K, well above the typical range for star-formation-dominated galaxies \citep[$\sim 20$--50\,K;][]{Casey2012, Magnelli2014}, suggesting that additional heating mechanisms, such as AGN activity may contribute to a non-negligible contamination to the far-infrared emission. Furthermore, if this assumption is incorrect then their true ${\rm SFR_{FIR}}$ will be lower than assumed (i.e.\ reducing the denominator in Equation~\ref{eq:fesc}), which would nominally increase the \fesc. However, the \lya\ and UV fluxes could also be boosted by the presence of an AGN, making it difficult to say if or how the \fesc\ would change if all effects were accounted for.

As shown in Figure~\ref{fig:escape_fraction}, we measure \fesc\ of 4--5\% for detected stacks (i.e.\ the more massive LAEs and those containing AGN), which is consistent with the global \fesc\ and previous studies at similar redshifts. For the stack of all AGN LAEs (Table~\ref{tab:stack_properties}), we assess the influence of AGN activity on \fesc\ by applying correction for AGN contamination to the ${\rm SFR_{FIR}}$ (see Section~\ref{sec:sedfit}). We find \fesc\,$\sim$5\% for the pure star-forming case. After correcting for AGN contamination, \fesc\ increases from 5\% to $\sim$8\% (QSO2; obscured case) and $\sim$10\% (QSO1; unobscured case). Thus, the inclusion of AGN contamination can increase \fesc\ by a factor of $\sim$1.6–2. Therefore, \fesc\ for LAEs containing AGN can be subject to systematic uncertainty due to the unconstrained AGN contribution to the infrared luminosity.

The lower limits for undetected stacks (i.e.\ average SC4K LAEs, non-AGN LAEs, and less massive galaxies) are higher than this, with lower limits that are calculated assuming a Sd-like infrared SED being significantly higher than those from the M82 SED template. The limits from the M82 template are generally similar to the evolution of the global \fesc\ and some previous studies \citep[e.g.][]{Erb2019, Roy2023}, and thus consistent with this evolution, though results for the Sd template are significantly higher, which hints that LAEs may have infrared SEDs that are warmer than that of Sd galaxies and more akin to M82-like systems. The $3\sigma$ lower limits on \fesc\ from the stack of the whole LAEs sample are  $\geq 21\%$ and $\geq 59\%$ assuming for the M82 and Sd templates, respectively. Both values are significantly higher than the global \fesc\ at comparable redshifts, which is consistent with the results from \citet{Haruka2015} and could indicate significant variation between LAEs (e.g.\ redshift evolution) such that a global average insufficiently represents the population. \citet{Matthee2021}, \citet{Haruka2015} and \citet{Melinder2023} contain galaxies with similarly high \fesc, which supports this possibility. Furthermore, this scatter can be attributed to a clumpy dust distribution within the ISM of LAEs, implying that variations in \fesc\ arise from dust geometry, as different fractions of \lya\ photons are able to escape through low-density channels.

In addition, all detected stacks covering a relatively narrow $z_{\text{median}}$ range ($2.99$ to $3.330$) are consistent with the results of \citep{Sobral2018}, which show an increase from $\approx 3.8\%$ at $z \approx 2.2$ to $\approx 15\%$ at $z \approx 6$, following the relation \fesc\ $\propto (1 + z)^{2.0 \pm 0.3}$. However, the choice of methodology may introduce systematic differences. By calculating \fesc\ directly from individual SFRs via equation~\ref{eq:fesc} rather than relying on global, model-dependent averages such as ionization efficiency and Case-B recombination; our method results in higher \fesc\ limits. On the other hand, the three individual detections falling below the SC4K trend indicate that local, source-by-source physical variations, such as complex dust geometry or gas kinematics, can deviate significantly from global, population-averaged evolutionary models. Deeper infrared studies of more individual LAEs are required to fully understand any such variation. 

Moreover, the detection of non-zero \fesc\ in these dusty, massive and AGN-hosting LAEs suggests that such systems can leak ionizing photons despite substantial dust attenuation. Since \fesc\ is often associated with ISM conditions that facilitate Lyman-continuum (LyC; ionizing photon) leakage \citep{Verhamme2015, Gazagnes2020, Begley2024}, the observed non-zero \fesc\ indicates the presence of low-column-density channels through which ionizing photons can escape. Although these galaxies lie at $z\sim2.2$--6 and therefore do not directly contribute to the epoch of reionization, they provide valuable analogues for understanding the mechanisms that enabled ionizing-photon escape in high-redshift galaxies for which estimation of LyC escape is extremely difficult due to the presence of neutral intergalactic medium \citep{Dijkstra2014b, Inoue2014, Becker2021b}. Our results potentially challenge the common assumption that dusty and massive galaxies are necessarily inefficient ionizing photon leakers \citep[e.g.,][]{Begley2022, Chisholm2022, Saldana-Lopez2022} and are more consistent with \citet{Casey2014b} and \citet{Casey2026}, who showed that DSFGs can exhibit blue $\beta_{\rm UV}$, implying that UV photons may escape even from heavily dust-obscured galaxies.

\subsection{IRX-$\beta$ Relation}
In many studies rest-frame UV/optical observations are used to measure UV luminosity and to derive parameters such as SFR \citep[e.g.][]{Shapley2003, Reddy2009, Oesch2010, Bouwens2015, Finkelstein2015}. However, these photons can be absorbed by dust and thus such measurements must be corrected for dust attenuation. Whilst UV/optical measurements such as the Balmer decrement (${\rm H{\alpha}/H{\beta}}$; e.g.\ \citealt{GarnBest2010, Reddy2015}) can be used, these do not account for regions that are entirely optically-thick at such wavelengths. Another option is to use the observed slope of the rest-frame UV emission, $\beta_{\rm UV}$, and to calibrate it using the correlation between $\beta_{\rm UV}$ and the infrared excess, ${\rm IRX = L_{IR} /L_{UV}}$ \citep{Meurer1999, Calzetti2000, Gordon2003, Reddy2018}. These empirical relations are based on local star-forming galaxies but hold for  high redshift LBGs albeit with higher scatter than locally \citep{Capak2015, Bouwens2016, Barisic2017, Bowler2018}. Measuring IRX for LAEs is more challenging, due to their faintness at far-infrared wavelengths, though \citet{Haruka2015} used their $3\sigma$ upper limits on \lfir\, derived from the MIPS and PACS stacks of 213 $z \simeq 2.18$ LAEs to determine upper limits on IRX. The PACS-based limit is shallow and consistent with \citet{Meurer1999, Calzetti2000} and \citet{Gordon2003}, while the MIPS-based limit is deeper. However, these limits together span a wide range, thus leaving IRX effectively unconstrained for LAEs.

\begin{figure}
\begin{center}
\includegraphics[scale=0.373]{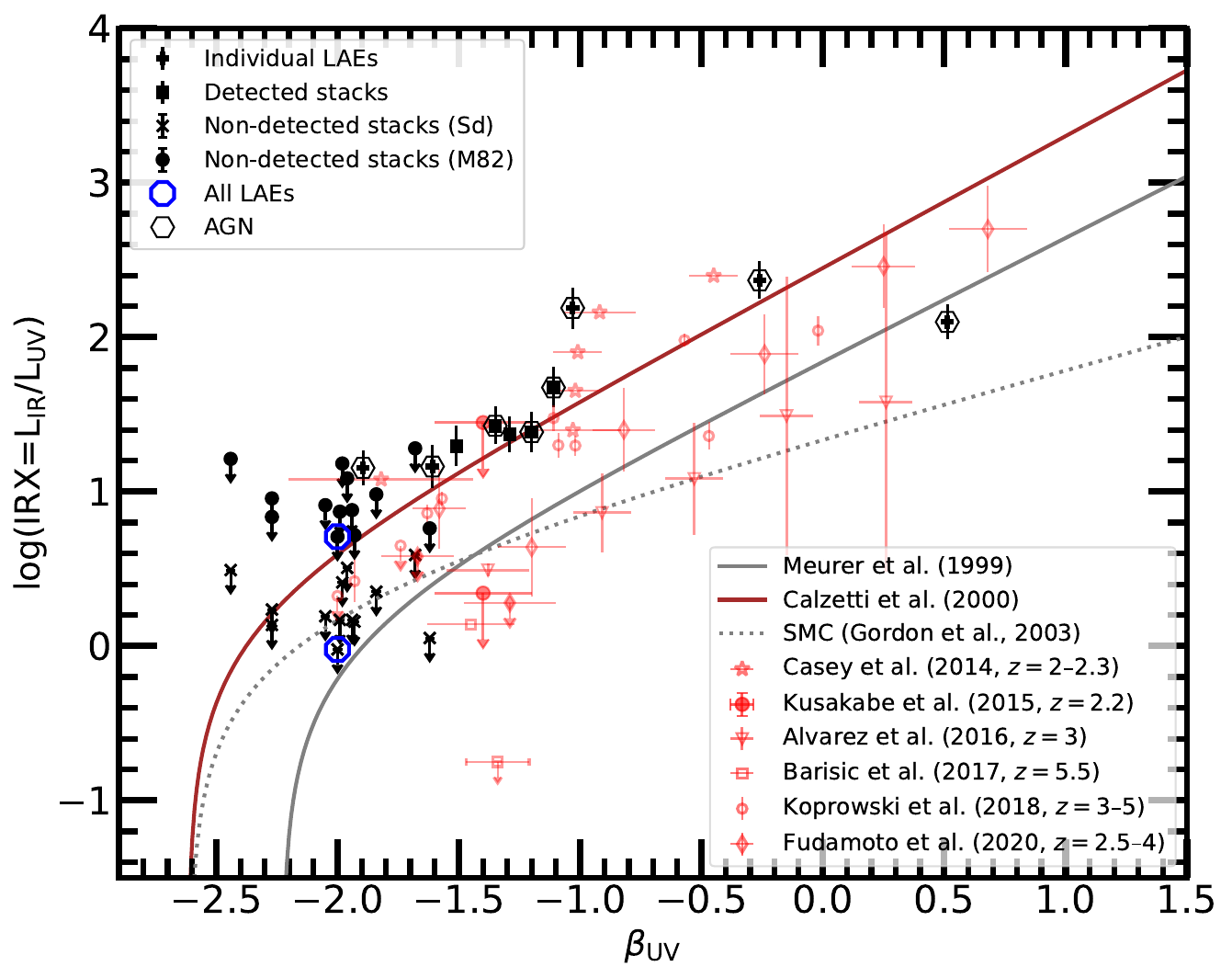}
\caption{The location of our individually-infrared detected LAEs and stacked subsets in the IRX-$\beta_{\rm UV}$ plane, compared with previous LAE \citep{Haruka2015}, DSFGs \citep{Casey2014b} and LBG \citep{Alvarez2016, Barisic2017, Koprowski2018, Fudamoto2020} studies and empirical relationships from \citet{Meurer1999, Calzetti2000, Gordon2003} with adjustments from \citet{Reddy2018}. For stacks that are not detected we show upper limits assuming both the Sd and M82 infrared SED templates (which encompass the full range of temperatures in the SWIRE template catalogue \citealt{Polletta2007}), and we highlight the stacks of all LAEs and those containing AGN. Overlapping points have been shifted by 0.1 in the x-direction for visibility. The detected LAEs and stacks typically have higher $\beta_{\rm UV}$ and IRX than the medians of the stacks that are not detected, consistent with the dust attenuation relations shown. The detected LAEs and stacks appear to contain dust that is more similar to the \citet{Calzetti2000} attenuation relation than the SMC or the \citet{Meurer1999}, however this may be attributed to the limited depth of single-dish observations, which preferentially detect the most massive systems and LAEs in which the dust is potentially AGN-heated.}
\label{fig:IRX_beta}
\end{center}
\end{figure}

We calculate IRX for our stacks and infrared-detected LAEs using the \lfir\, values calculated in Sections~\ref{sec:sedfit} and \ref{sec:fesc}. We use the $\beta_{\rm UV}$ values from the SED fits in \citet{Santos2020} and calculate the median $\beta_{\rm UV}$ for stacked samples. Figure~\ref{fig:IRX_beta} compares our IRX-$\beta_{\rm UV}$ measurements with the empirical relationships from \citet{Calzetti2000} (i.e.\ local star-burst galaxies with relatively higher intrinsic $\beta_{\rm UV, 0}$ and metallicity), \citet{Meurer1999} (i.e.\ starburst galaxies with lower intrinsic $\beta_{\rm UV, 0}$ and moderate metallicity) and \citet{Gordon2003} (i.e.\ SMC-like dust with lower metallicity). We use the updates from \citet{Reddy2018} for the \citet{Calzetti2000} and \citet{Gordon2003} relations, which assumes a $\beta_{\rm UV, 0}=-2.62$, metallicity of $0.14Z_{\odot}$, constant star-formation, and a stellar population age of 100 Myr. We also compare  with the data for LAEs from \citet{Haruka2015}, LBGs \citep{Alvarez2016, Barisic2017, Koprowski2018, Fudamoto2020} and with DSFGs \citep{Casey2014b} at similar redshifts. Our results are consistent with the upper limits for LAEs from stacked PACS data from  \citet{Haruka2015}, though most of our measurements are significantly higher than the MIPS 24\,\mic\ stacked results presented in \citet{Haruka2015}: this is likely due to the uncertainties associated with extrapolating from 24\,\mic\ to the integrated far-infrared luminosity. All except one of our detected stacks (the X-ray AGN) and individually infrared-detected LAEs (SC4K-IA484-28746) lie on or below the \citet{Calzetti2000} attenuation curve, suggesting that these systems have similar dust to local starburst galaxies. The far-infrared detected LAEs and stacks (i.e.\ the most massive systems and those containing AGN) have significantly higher IRX (by $\sim 0.7$\,dex) than the LBGs from \citet{Alvarez2016} and \citet{Barisic2017} and they are also slightly systematically higher than the LBGs studied by \citet{Koprowski2018} and \citet{Fudamoto2020}. They are more similar to the DSFGs at $z=2$--2.3 \citep{Casey2014b}, which show relatively blue $\beta_{\rm UV}$ despite being dusty. This is likely because their UV emission is dominated by recent star formation, while the dust is produced by short-lived, massive O- and B-type stars. 

We also compare our obscured star formation fraction (i.e., $\sim$91--99\% for far-infrared detected and $\lesssim$79\% for average LAE) with the expectations based on \citet{Whitaker2017}, who showed that more massive galaxies are expected to contain a higher fraction of obscured star formation. We find that our results are consistent with the trend reported by \citet{Whitaker2017}. Collectively, these results suggest that the most massive LAEs may have larger dust reservoirs, with AGN activity potentially enhancing dust heating and far-infrared luminosity. Alternatively, the observed elevated IRX may be explained by clumpy or uneven dust geometry, or by a dust screen with holes in LAEs. In such scenarios, patchy dust can allow some UV light to escape while dense regions can reprocess a significant fraction of the radiation into the far-infrared, suggesting that \lya\ and dust emission can originate from same or different regions in LAEs, as proposed by \citet{Oteo2012} and \citet{Popping2017}.

For the stacks that are not-detected in the infrared we show the $3\sigma$ upper limits on IRX from both the Sd and M82 SED templates, which encompass the range of infrared SED temperatures \citep{Polletta2007}. Limits that are calculated assuming the Sd-like SED template are significantly lower than those from the M82 template and would suggest that LAEs are more likely to follow a SMC- or \cite{Meurer1999} attenuation curve. However, the limits determined from the M82 SED (which gives more typical \fesc\ values; Figure~\ref{fig:escape_fraction}) are consistent with the \citet{Calzetti2000} relation. Thus, even with a catalogue of $\sim4000$ LAEs and stacking some of the deepest \herschel\ and SCUBA-2 data available we are still unable to measure IRX for LAEs with the bluest UV slopes (i.e.\ $\beta_{\rm UV}\lesssim-1.5$). Deeper measurements are required to determine whether lower-mass LAEs have similar IRX-$\beta_{\rm UV}$ properties to the higher mass population (which typically have redder UV slopes), though our limits are consistent with this interpretation.

We also investigate the dependence of \lfir\ on \lya\ luminosity and stellar mass. We do not observe any trend between \lfir\ and \lya\ luminosity, however we find a tentative trend between \lfir\ and stellar mass, where more massive LAEs tend to show higher far-infrared luminosities, which is consistent with the findings from \citet{Whitaker2017}. Since \lfir\ is widely used as a tracer of SFR and is known to follow the galaxy main sequence \citep[e.g.][]{Whitaker2012, Lee2013, Schreiber2015A&A}, it naturally correlates with stellar mass. Moreover, previous work on the same sample of LAEs from \citet{Santos2021}, reports a correlation between \lya\ luminosity and stellar mass, this suggests that both stellar mass and \lya\ emission may be indirectly linked through ongoing star formation, which both produces \lya\ photons and drives dust enrichment.

\section{Conclusions} \label{Sec:5}
We have analysed the average dust emission and far-infrared properties of 3674 LAEs together with five individually far-infrared detected LAEs at $z = 2.2$--6 in the SC4K survey of the COSMOS field \citep{Sobral2018}. We used publicly available \herschel\ PACS and SPIRE data at 100, 160, 250, 350 and 500\,\mic\ \citep{Lutz2011, Oliver2012, Shirley2021} and SCUBA-2 850\,\mic\ observations from the S2COSMOS survey \citep{Simpson2019}. 
We identified five LAEs that are sufficiently bright to be detected in the SCUBA-2 850\,\mic\ data, and used stacking to probe the submillimetre emission from the remaining LAEs. We performed far-infrared SED fitting, using MCIRSED where possible (i.e.\ for systems with detections at $\ge2$ far-infrared wavelengths) and SWIRE galaxy templates otherwise. We used these results to measure IRX-$\beta_{\rm UV}$ and \fesc\ for 
the LAEs that were individually-detected in the submillimetre and for median stacks of subsets of LAEs. 
Our main results are:
\begin{enumerate}
\item The five LAEs that are bright enough to be individually detected in the far-infrared without stacking, all contain radio and/or X-ray AGN and have higher average \lya\ luminosity than the SC4K LAEs as a whole. 
The origin of \lya, continuum UV, and far-infrared emission from these LAEs is uncertain and may include contributions from the AGN. 

\item A stack of all the SC4K LAEs probes down to $\sigma=0.03$\,mJy at 850\,\mic\ but this median stack of LAEs remains undetected. We also stacked subsets of LAEs in bins of redshift, \lya\ luminosity, stellar mass, and distinguished by the presence of AGN. In these stacks dust emission was detected from the stacks of all AGN, X-ray-selected AGN, radio-selected AGN and the most massive LAEs ($\mstar =10^{10-12}$\,\msun).

\item Whilst many of the most massive LAEs contain AGN, we also excluded the AGN and stacked the remaining $\mstar =10^{10\--12}$\,\msun\ galaxies, showing that these non-AGN but massive LAEs also produce significant dust emission. Thus, the submillimetre emission from LAEs is separately correlated with both stellar mass and AGN activity. Deeper far-infrared data are required to detect dust emission from the lower mass, fainter LAEs.

\item We measured $\fesc=1$--7\% for the five  LAEs that were individually-detected in the far-infrared data, and the subsets of LAEs that were detected in the stacks have $\fesc=4$--5\%. These are similar to previous studies and consistent with the global evolution of \fesc\ at similar redshifts. For AGN–hosting LAEs, the inclusion of potential AGN contamination increases \fesc\ by a factor of $\approx$\,1.6–2.

\item For far-infrared undetected stacks, including the stack of entire population of LAEs, we explored the effects of different infrared SED templates and found that the \fesc\ that were derived from assuming a Sd-like SED template give $\fesc\geq 33\%$ ($3\sigma$), which is significantly higher than the global \fesc\ at comparable redshifts. Limits derived using a M82 SED template are closer to the evolution expected from \citet{Hayes2011}, which suggests that LAEs likely have warmer dust than local Sd-like galaxies. The stack of the all SC4K LAEs has a higher limit on \fesc\ than expected from the global evolution and may indicate a significant variation or evolution within the population. Moreover, the observed variation and scatter in \fesc\ (1 to $\geq21\%$) can be attributed to a clumpy ISM dust distribution, in which different fractions of \lya\ photons are able to escape depending on the local dust structure.

\item The far-infrared detected LAEs and median stacks all have redder UV slopes ($\beta_{\rm UV}$) than those that were not detected. They also tend to have higher IRX than LBGs, though our results are consistent with the \citet{Calzetti2000} dust attenuation relation, however, the current detections are biased toward systems with higher stellar masses and AGN activity. Therefore, deeper observations are required to include low-mass and far-infrared–faint LAEs to better constrain this relation. The dust in these galaxies is likely characterized by a clumpy or patchy distribution, or by dust screens with holes that facilitate \lya\ escape. This geometry introduces strong variations in the effective optical depth along different sightlines, thereby producing increased scatter in the observed escape fraction. At the same time, a significant fraction of UV photons can still be absorbed and reprocessed by dust, leading to elevated infrared emission and thus potentially high IRX values despite relatively blue UV slopes.

\end{enumerate}
A tentative trend between far-infrared inferred properties and stellar mass and a correlation between \lya\ luminosity and stellar mass for the same sample of LAEs reported in \citet{Santos2021} suggest that both stellar mass and \lya\ emission are likely to be connected through ongoing star formation which can  produce both \lya\ photons and dust content. Moreover, non-zero observed \lya\ emission from these dusty, massive and AGN-hosting LAEs suggests ionizing photons might escape despite the presence of dust. Though at $z\leq$\,6, they serve as analogues for high-redshift LyC escape and challenge the idea that dusty massive galaxies are inefficient LyC leakers. Deeper far-infrared data, particularly data that are able to probe a variety of individual LAEs, are required to understand the diversity in this population and confirm the effect of stellar mass and AGN on their dust emission. Our study has shown the depths required, which can currently only be obtained with interferometric instruments such as ALMA due to the confusion limit of single-dish observations. 

\section*{Acknowledgements}

RR acknowledges a National Overseas Scholarship from the Ministry of Tribal Affairs, Government of India and the ERC synergy grant 101166930 (RECAP). JLW acknowledges support from an STFC Ernest Rutherford Fellowship (ST/P004784/2). We also acknowledge the High-End Computing (HEC) cluster of Lancaster University. TC acknowledges support from STFC under grant ST/X00127X/1. Additionally, the authors acknowledge the values of publicly available programming language {\sc Python} and libraries {\sc Astropy} \citep{Robitaille2013, Whelan2018}, {\sc Matplotlib} \citep{Matplotlib},  {\sc NumPy} \citep{Numpy} and {\sc SciPy} \citep{Scipy2020}. This work also makes use of {\sc TOPCAT} \citep{Topcat}. 
 
 \section*{Data Availability}

All data used in this study are publicly available. LAEs catalogue are from SC4K \citep{Sobral2018}. Far-infrared observations include \textit{Herschel}/PACS 100 and 160\,\mic\ from the PACS Evolutionary Probe (PEP; \citealt{Lutz2011}) and \textit{Herschel}/SPIRE 250, 350, and 500\mic\ from HELP \citep{Shirley2021}. JCMT/SCUBA-2 850\,\mic\ data are from S2COSMOS \citep{Simpson2019}. All datasets are accessible as described in the references.


\bibliographystyle{mnras}
\bibliography{Ref_v2} 
\appendix
\section{}
\begin{figure*}
\begin{center}
\includegraphics[scale=0.6]{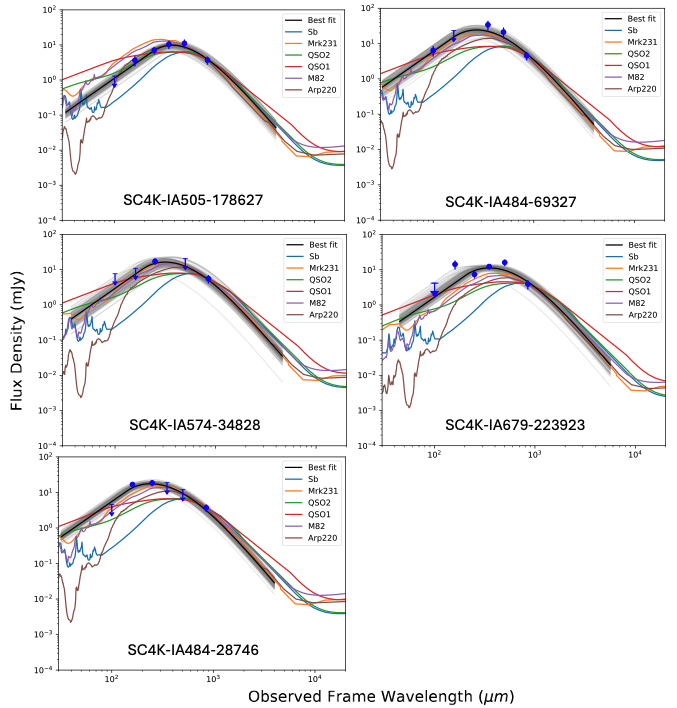}
\caption{Observed-frame SEDs of five individually detected LAEs listed in Table~\ref{tab:six_LAEs_flux1} and shown in Figure~\ref{fig:six_LAEs}. Observed fluxes and upper limits at 100, 160, 250, 350, 500 and 850\,\mic\ are shown with blue symbols. The MCIRSED best fit is presented as solid black curve, while the gray lines represent 100 other probable SED models. The coloured curves correspond to SWIRE templates \citep{Polletta2007}, scaled to the 850\,\mic\ flux density, which is also the deepest observation. 
The apparent inconsistent flux densities of the LAE (SC4K-IA679-223923) may stem from super-deblending \citep{Jin2018}, potentially caused by over-subtraction and/or residual blending. As we have scaled SWIRE templates to the deepest observation, some templates lie within $\sim$1$\sigma$ of the best fit, so the derived \lfir\ remains broadly consistent within the errors.}
\label{fig:SED_Individual}
\end{center}
\end{figure*}

\bsp	
\label{lastpage}
\end{document}